\documentclass[journal]{IEEEtran}
\usepackage{cite}
\usepackage{url}

\usepackage{color,soul}
\usepackage[T1]{fontenc}
\usepackage{tikz,forest}
\ifCLASSINFOpdf
\usepackage{graphicx}
\usepackage{tabularx}
\usepackage{amssymb}
\usepackage{float}
\usepackage{enumitem}
\graphicspath{{./pdf/}{./jpeg/}}
\else
\fi

\usepackage{algorithm}
\usepackage[noend]{algpseudocode}

\let\emptyset\varnothing

\makeatletter
\def\BState{\State\hskip-\ALG@thistlm}
\makeatother

\usepackage{array}

\ifCLASSOPTIONcompsoc
  \usepackage[caption=false,font=normalsize,labelfont=sf,textfont=sf]{subfig}
\else
  \usepackage[caption=false,font=footnotesize]{subfig}
\fi
\usepackage[cmex10]{amsmath}
\usetikzlibrary{calc}
\usetikzlibrary{positioning}
\usetikzlibrary{decorations.text}
\usetikzlibrary{arrows.meta}
\usepackage{multirow}

	\usepackage[absolute,showboxes]{textpos}

\TPMargin{5pt}

\begin{document}
%
\title{MORPH: An Adaptive Framework for Efficient and Byzantine Fault-Tolerant SDN Control Plane}

\author{Ermin~Sakic,~\IEEEmembership{Student Member,~IEEE,} \\ Nemanja~\DJ{}eri\'c{},~\IEEEmembership{Student Member,~IEEE,}
	\\ Wolfgang~Kellerer,~\IEEEmembership{Senior Member,~IEEE}%
		\thanks{E. Sakic is with the Department
		of Electrical Engineering and Information Technology, University of Technology Munich, Germany;
		and Corporate Technology, Siemens AG, Munich, Germany, E-Mail: (ermin.sakic@\{tum.de, siemens.com\}).}%
		\thanks{N. \DJ{}eri\'c{} is with the Department of Electrical Engineering and Information Technology, 
		University of Technology Munich, Germany, E-Mail: (nemanja.deric@tum.de).}
		\thanks{W. Kellerer is with the Department of Electrical Engineering and Information Technology, 
		University of Technology Munich, Germany, E-Mail: (wolfgang.kellerer@tum.de).}
}
\newcommand{\todo}[1]{\textcolor{red}{#1}\PackageWarning{TODO:}{#1!}}



%


\maketitle

\begin{abstract}
	Current approaches to tackling the single point of failure in SDN entail a distributed operation of SDN controller instances. Their state synchronization process is reliant on the assumption of a correct decision-making in the controllers. Successful introduction of SDN in the critical infrastructure networks also requires catering to the issue of \emph{unavailable}, \emph{unreliable} (e.g. buggy) and \emph{malicious} controller failures. We propose MORPH, a framework tolerant to unavailability and Byzantine failures, that distinguishes and localizes faulty controller instances and appropriately reconfigures the control plane. Our controller-switch connection assignment leverages the awareness of the source of failure to optimize the number of active controllers and minimize the controller and switch reconfiguration delays. The proposed re-assignment executes dynamically after each successful failure identification. We require $2F_M+F_A+1$ controllers to tolerate $F_M$ malicious and $F_A$ availability-induced failures. After a successful detection of $F_M$ malicious controllers, MORPH reconfigures the control plane to require a \emph{single} controller message to forward the system state. Next, we outline and present a solution to the practical correctness issues related to the \emph{statefulness} of the distributed SDN controller applications, previously ignored in the literature. We base our performance analysis on a resource-aware routing application, deployed in 
	an emulated testbed comprising up to $16$ controllers and up to $34$ switches, so to tolerate up to 5 unique Byzantine and additional 5 availability-induced controller failures (a total of $10$ unique controller failures). We quantify and highlight the dynamic decrease in the packet and CPU load and the response time after each successful failure detection. 
\end{abstract}


%
\IEEEpeerreviewmaketitle

\section{Introduction}

In a Software Defined Network (SDN), switches and routers rely on a centralized SDN controller to provide the necessary configurations for path finding and resource configuration. In a \emph{single-controller} SDN, the controller represents the single point of failure. In case of a disabled or compromised controller, the SDN control plane becomes inoperable. The user faces the possibility of losing control over the network equipment and hence risks the unavailability of the data plane for new services. Indeed, the availability of the network control and management functions is of paramount importance in industrial networks where critical network failures may lead to disruption of applications, potentially leading to safety and regulatory issues \cite{race, virtuwind, huang2016real}. Hence, for the purpose of resilience, a multitude of strategies for deploying parallel SDN controller instances were proposed in the recent literature \cite{bft-smart, hyperflow, ravana, odl, berde2014onos}. 

In state-of-the-art distributed SDN controller implementations \cite{odl, berde2014onos}, a resilient control plane is established by a controller-switch role-assignment procedure. In these architectures, for each switch, exactly one controller is assigned the role \emph{PRIMARY}, and multiple backup controllers are assigned the role \emph{SECONDARY}. Hence, in the case of a critical failure in the PRIMARY controller, a SECONDARY controller can take over the failed PRIMARY controller's switches. When handling external requests such as switch events or client requests, only a single controller is declared the PRIMARY controller for a node. A good example is the addition of a new flow service in the network. Here, a PRIMARY controller may need to handle the request in the following manner:
\begin{enumerate}
	\item If the PRIMARY controller receives the client request, the task (e.g., path finding) is executed locally. 
	\item If a SECONDARY controller receives the client request, the request is proxied by the SECONDARY to the PRIMARY controller, which then proceeds with request handling. If the PRIMARY controller fails, a new PRIMARY that handles the client request, is re-elected \cite{ongaro2014search, lamport2001paxos}.
\end{enumerate}

However, an occurrence of a Byzantine failure of the PRIMARY controller (i.e. because of a  corrupted \cite{race}, manipulated \cite{yan2016software}, or buggy \cite{vizarreta2017characterization} internal state), may lead to incorrect computation decisions in the controller logic. With Byzantine failures, there is incomplete information on whether the controller has truly failed. Thus, the controller may appear as both failed and correct to the failure-detection systems, presenting different symptoms to different observers (i.e. switches). Additionally, a malicious adversary may interfere with or modify the controller logic for the purpose of taking control of the network or re-routing the network traffic \cite{yan2016software}. 

On the other hand, availability-related controller failures lead to the controllers becoming inactive. In contrast to Byzantine failures, such controllers do not actively perform malicious actions nor do they try to hide their real status. Following an availability failure they stop reacting to accepting client requests and eventually time out. Availability types of failures are detected using failure detectors and not by means of a semantic message comparison. Existing state-of-the art works\cite{mohan2017primary, li2014byzantine}, however, do not distinguish Byzantine from availability-related failure sources, such as the failure of the underlying hardware, hypervisor or data plane links that interconnect the controllers. They consider the availability-related failures a subset of Byzantine failures and thus tend to overprovision the required number of controller instances.

The approach presented henceforth leverages a successful discovery of controller failures \emph{and} the cause of the failures, in order to optimally adapt the number of deployed controllers as well as the controller-switch assignments during runtime. Our approach realizes a low-overhead control plane operation, while protecting from mentioned types of faults at all times.

\subsection{Our Contribution}
We handle the Byzantine faults by modeling the controller instances of the distributed SDN control plane as a set of Replicated State Machines (RSMs). MORPH proposes a dynamic re-association of the controller-switch connections after detecting a fault injected by a malicious adversary or an availability-related failure of an SDN controller instance. Depending on the network operator's preference, instead of contacting just the PRIMARY controllers, the edge switches may also forward the client requests to both PRIMARY and SECONDARY controllers (i.e. using an OpenFlow \emph{packet-in} message \cite{mckeown2008openflow}) in order to request additional responses, computed by the isolated controller instances. The switch then collects the different responses and evaluates them for inconsistencies. Since the controllers are modeled as RSMs, each \emph{correct} controller is expected to compute the exact same response for any given input request. The SECONDARY controllers may forward their configuration messages either \emph{proactively}, or \emph{reactively}, when inconsistencies in the configuration messages are detected in the switch. Finally, each switch deduces the correct decision locally, based on the minimum number of required consistent (matching) controller configuration messages. 

In MORPH, we introduce three architectural components: i) \emph{REASSIGNER} - detects the controller failures, differentiates the types of failures and reassigns the controller-switch connections on a detected failure; ii) \emph{S-COMPARATOR} - the switch component that compares the controller responses for the purpose of detection of inconsistent configurations; iii) \emph{C-COMPARATOR} - the controller agent which compares controller-to-controller messages in order to identify inconsistent state synchronization messages. The comparators additionally apply the controller-switch assignment lists provided by the REASSIGNER. The REASSIGNER recomputes the controller-switch lists, with the objective of minimizing the amount of considered controllers in deduction of new configuration and thus reducing the worst-case waiting period required to confirm any new switch configuration. 

We introduce the dynamic controller-switch reassignment procedure that optimally assigns the controllers to switches and makes strict guarantees for the QoS constraints, such as the maximum controller-to-switch and controller-to-controller delays and the controller processing capacity. MORPH solves the controller-switch assignment problem during online operation using an Integer Linear Program (ILP) formulation. As a consequence, our switch agent applies the internal reconfigurations after receiving a \emph{maximum} of $F_M+1$  consistent/matching controller messages, thus minimizing the response time compared to \cite{li2014byzantine} and \cite{mohan2017primary} which require a \emph{minimum} of $3F+1$ and $F+1$ consistent messages to forward the switch state at any point in time, respectively. In MORPH, $F_M$ denotes the maximum number of tolerated Byzantine failures, whereas $F_A$ represents the number of maximum tolerated unavailability-induced failures. In common state-of-the-art literature, this differentiation is not done, hence, there $F$ denotes the number of tolerated combined Byzantine and availability-induced failures. The switches in MORPH require a minimum of \emph{one} controller message to apply the internal reconfigurations, after $F_M$ malicious controllers are detected in the system.

In our evaluation, we conclude that the dynamic reassignment of controller-switch connections results in a considerable performance improvement in: i) the observed best- and worst-case response time in controller-to-switch and controller-to-controller communication; ii) a decrease in generated packet load and iii) a decrease in experienced CPU load. 

The performance analysis of our implementation comprises a varied set of maximum tolerated controller failures. Measurements were executed on two well-known large- and medium-sized service provider and data-center topologies. The emulated testbed comprised distributed MORPH processes and Open vSwitch instances and is thus a realistic approximation of the expected actual performance. 

We structure the paper as follows. Important terminology is introduced in Sec. \ref{terminology}. The state-of-the-art literature is presented in Sec. \ref{relatedwork}. Sec. \ref{soaissues} highlights the practical correctness issues related to the relevant state-of-the-art BFT approaches \cite{li2014byzantine, mohan2017primary}. In Sec. \ref{systemmodel} we introduce the MORPH architecture.  Sec. \ref{algorithm} presents  the algorithms implemented by MORPH. Sec. \ref{ilp} details the ILP formulation for the dynamic controller-switch assignment procedure. Sec. \ref{eval} outlines our evaluation methodology. We present the evaluation results in Sec. \ref{results}. Finally, Sec. \ref{conclusion} concludes this paper.

\section{Terminology}
\label{terminology}

To clarify the issues with the existing approaches, we introduce the following important concepts:
\begin{itemize}[leftmargin=*]
	\item A \emph{malicious controller} is an active controller in possession of a malicious adversary. It is able to compute correct and incorrect responses to the client requests and propose the according configurations to the switches. It may disguise itself as a \emph{correct} controller for an arbitrary period of time. In our model, unreliable controllers that compute an incorrect result as a consequence of a buggy internal state are also considered "malicious". Until suspected, all malicious controllers are considered \emph{correct}.
	\item A \emph{correct controller} is an active controller instance which is not \emph{malicious} and which actively participates in the cluster of MORPH controllers.
	\item An \emph{unavailable controller} is a disabled controller which is not active as a result of a hardware/software failure. It is unknown if an unavailable controller was either \emph{correct} or \emph{malicious} at some point in time before its failure.
	\item An \emph{incorrect controller} is a controller that is either malicious or unavailable. 
	\item \emph{State-independent (SIA)} SDN applications: We refer to SDN applications that do not require the inter-controller state synchronization as \emph{state-independent applications} (\emph{SIA}), e.g. shortest path routing without resource guarantees, controller-supported MAC-learning, load balancing etc. 
	\item \emph{State-dependent (SDA)} SDN applications: For reservation-based applications, state synchronization between the controllers is necessary to generate correct new decisions for requests which rely on decisions made in the past (i.e. the \emph{causality} property). We refer to these types of applications as \emph{state-dependent applications} (\emph{SDA}), e.g. reservation-based path finding and resource management \cite{guck2016function}, flow scheduling, optimal and reservation-aware load balancing \cite{levin2012logically} etc.
\end{itemize}

\section{Related work}
\label{relatedwork}
In this section, we present the related work in the context of Byzantine Fault Tolerance (BFT) research in SDN.

The prominent open-source SDN platforms such as OpenDaylight\cite{odl} and ONOS\cite{berde2014onos} support no means of identifying Byzantine failures in the controller cluster. However, they do implement measures to support controller fail-over in the face of availability-induced failures. More specifically, ONOS and OpenDaylight allow for replication of the controller state in a \emph{strong} and \emph{eventually} consistent manner \cite{sakicadaptive, sakicresponse}, so to provide for availability of the data-store knowledge in the SECONDARY controllers, after a failure of a PRIMARY. Their strong consistent data-store is based on RAFT consensus algorithm \cite{howard2015raft, ongaro2014search, suh2016performance}. RAFT is, however, susceptible to Byzantine failures, as only a single leader is elected at a time by design. The RAFT \emph{leader} processes and broadcasts any new data-store updates to the \emph{followers} before an update may be considered committed. This introduces multiple attack vectors, e.g.: i) the adversary may generate malicious state updates in each follower if the leader controller is in the possession of a malicious adversary; ii) inconsistencies in the state view \cite{levin2012logically} of master and followers if adversary takes over the RAFT follower; iii) incorrectness issues, e.g., a continuous non-convergence of controller cluster leadership \cite{zhang2017raft}.

The definition of the problem of malicious SDN controller instances, including an initial solution draft, was proposed in \cite{li2014byzantine}. The authors discuss the requirement of a minimum assignment of \emph{3F+1} controllers to each switch \emph{at all times}, so to tolerate a total number of \emph{F} Byzantine failures. \emph{3F+1} matching messages are required during the \emph{agreement} and \emph{2F+1} during the \emph{execution} phase in order to reach the majority and consensus on the correct response after the comparison of the computed configuration outputs \cite{yin}. 

	\emph{BFT-SMaRt} \cite{bft-smart} proposes a strong consistent framework for supporting Byzantine- and availability-induced failures. The authors abstract away the notion of a "failure" to consider controllers \emph{failed}, if either the controller process crashes and never recovers or the controller keeps infinitely crashing and recovering. Their follow-up work \cite{botelho2016design} applies the \emph{Bft-SMaRt} in order to improve the data-store replication performance in a strong consistent SDN. They evaluate the workload generated by real SDN applications as they interact with the data-store.

	Both works do not consider the issue of malicious controller-to-controller synchronization, where a controller may initiate malicious state synchronization procedure and thus commit malicious database changes in the correct controllers. Furthermore, they do not cover for the aspects of an adaptive controller-switch connection reassignment procedure nor do they distinguish and leverage the difference between the Byzantine and availability-related controller failures.	

	A recent proposal for a BFT-enabled SDN \cite{mohan2017primary} advocates the usage of a total of \emph{2F+1} instead of \emph{3F+1} controller instances. The authors propose the collection of \emph{F+1} PRIMARY controller configuration messages at each switch (generated by their PRIMARY controllers), and a delayed request to the remaining \emph{F} SECONDARY instances if an inconsistency is detected in the switch. The configuration message is flagged as correct only after a minimum of \emph{F+1} matching configurations were received in the switch. The authors do not consider the source of failure nor the effect of their design on the controller state synchronization process.

\section{Issues with the Current Designs}
\label{soaissues}
Apart from the inefficiency issues related to the unnecessary control plane packet load during normal (non-faulty) operation in \cite{li2014byzantine} and \cite{mohan2017primary}, we have identified multiple correctness-related issues not addressed by these works: 

\emph{Inadequate constellation of controllers}: The \emph{2F+1} mode of configuration leads to an incorrect decision-making when \emph{correct} instances become \emph{unavailable} due to a controller-to-switch link, controller-to-controller link or a controller instance failure because of an availability-related issue, \emph{previous} to the identification of \emph{F} of the remaining \emph{F+1} active instances as \emph{malicious}. Namely, the switch may incorrectly interpret the majority of Byzantine messages as correct. Depending on the design choice, after receiving \emph{F+1} inconsistent (non-matching) controller replies, the switch may decide to not accept any future configuration sent by the remaining active controllers or worse, accept the proposed malicious configuration (the majority configuration) as the valid configuration. Either design results in an unavailability and/or incorrectness of the system after \emph{one} correct replica has failed/or is unreachable because of the unavailability preceding the activation of the malicious controller replicas. Thus, both the number \emph{and} the order and type of failures matter for the correct failure localization. Our approach requires an initial deployment and switch-controller assignment of $2F_M+F_A+1$ controllers to tolerate $F_M$ malicious adversary- and $F_A$ availability-induced failures (assuming this the total request processing capacity constraint holds). During the runtime, we dynamically adapt the remaining number of assigned controllers according to the number and type of detected controller failure.

\emph{Insecure controller-to-controller channel}: The existing approaches do not address the issue of securing the controller-to-controller communication. For SDA applications, letting a malicious controller compute a correct operation and distribute the Byzantine state changes (e.g. reservations) to the remaining controllers may result in committing those malicious state changes into the data-stores of the correct controller instances. The issue is exacerbated when reservation values (i.e. the bandwidth and buffer reservations, flow table occupancies etc.) are exchanged between controller instances for the purpose of enabling resilience for mission-critical communication services. Namely, the global controller decisions potentially stretch across switches under control of different administrative controller clusters, with each active controller requiring a consistent global state to achieve optimal decision-making.

\begin{figure}[htb]
	\centering
	\includegraphics[width=0.5\textwidth]{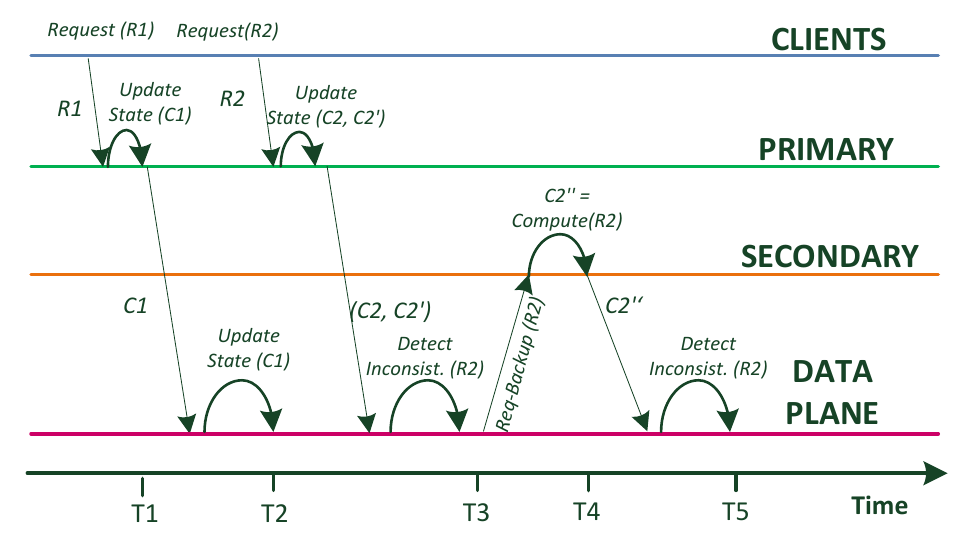}
	\caption{State-dependent applications: Issue of making incorrect decisions or non-convergence on the new switch configuration when fetching SECONDARY controller responses.}
	\label{fig:inconsistency}
\end{figure}

\emph{Distinguishing false positives for SDA}: Requesting the computation of $F$ SECONDARY controller configuration responses in addition to the collected \emph{F+1} PRIMARY responses, without reasoning about the state awareness, may lead to an incorrect identification of malicious controllers. 

A problem scenario is depicted in Fig. \ref{fig:inconsistency}. According to \cite{mohan2017primary}, given an input request $R1$, PRIMARY controllers compute a consistent switch configuration $C1$ at time $T1$. The switch accepts the configuration $C1$ only if \emph{F+1} PRIMARY replicas have computed a consistent result. As this is the case for $R1$, the switch commits the new configuration $C1$ at time $T2$. Similarly, the PRIMARY controllers update their controller state according to the configuration $C1$. Let the client request $R2$ arrive at the PRIMARY controllers at time $T2$. The PRIMARY controllers handle the request and forward the computed correct and Byzantine configurations $C2$ and $C2'$, respectively, to the switches. The configuration is inconsistent, hence the switches decide to request a recomputation of the result for request $R2$ at the SECONDARY controllers at time $T3$. Since the SECONDARY controllers have never observed the updated view of the internal data-store as a result of the changes related to $C1$, they compute a configuration $C2''$ at $T4$ that differs from the correct configuration $C2$ computed at the correct PRIMARY controllers at $T2$. Thus, the switches are unable to distinguish correct from incorrect controllers at $T5$ as \emph{F+1} consistent results may not be deducible. More importantly, the switches may mistake correct for malicious controllers (false positives). This is realistic in the scenario where the majority of PRIMARY controllers assigned to the affected switch are malicious.


\section{MORPH System Model}
\label{systemmodel}

We next introduce our system model and architecture in more detail and explain the novel system elements. As our starting point, we consider a typical SDN architecture where the data plane (i.e. networking elements/switches) and control plane are separated \cite{mckeown2008openflow}. We extend it with additional functional elements to enable a BFT and unavailability-tolerant system operation. The control plane communication between the switches and controllers (S2C, C2S) and between controllers (C2C) is realized via an in-band control channel \cite{schiff2016band}. Furthermore, all the communication between different elements (i.e. REASSIGNER, C-COMPARATOR and S-COMPARATOR) is assumed to be signed \cite{paladi2016trusdn}. Thus \emph{i) message forging is assumed impossible} and \emph{ii) message integrity is ensured in the data plane during normal operation}. In order to prevent a faulty replica from impersonating a correct replica, correct replicas can authenticate each message using MAC authentication \cite{castro1999practical,eischer2017}.

\subsection{Design Goals}
\label{designgoals}

The proposed architecture is designed to accomplish two major objectives, \emph{i) data plane protection} from incorrect controllers and \emph{ii) re-adaptation to the optimal network configuration} w.r.t. controller-to-switch connection assignments based on the present number of correct and incorrect controllers.

\emph{i) Data plane protection:} Even if there are up to $F_M$ malicious controllers and $F_A$ failed controllers in the network, the data plane networking elements must \emph{never} be incorrectly reconfigured or tampered with by the incorrect controllers. MORPH realizes this protection by leveraging \emph{replicated computation} of control plane decisions and their transmission from multiple controllers to the target switches. Hence, each switch is connected to multiple controllers at the same time, and only accepts and proceeds to commit the reconfiguration requests if a sufficient number of matching messages for reaching the \emph{correct consensus} was received. If we consider a scenario where $F_M$ malicious controllers send their reconfiguration requests to the networking elements, it becomes clear that we need at least $F_M+1$ correct reconfiguration requests in order to enable the controllers to reach a \emph{correct consensus} and for the switches to distinguish a correct reconfiguration request from an \emph{inconsistent} message set. For the detection of an unavailability-induced controller failure, we deploy a failure detector \cite{phi-accrual, hayashibara2002failure} that reliably identifies the \emph{fail-silent} controllers \cite{cachin2011introduction}. Hence, in the case of a failure of $F_A$ controllers, we only require one additional backup instance (i.e. \emph{$F_A+1$}) to achieve a correct control plane protection. Therefore, in order to protect the data plane from $F_M$ malicious controllers and $F_A$ failed controllers each switch has to be \emph{assigned} to at least $|\mathcal{C}|=2F_M + F_A + 1$ controllers. Granted, a switch will accept the new reconfiguration request already after receiving $F_M+1$ matching messages from the controllers.

Additionally, individual switch failures may cause packet loss and thus an unsuccesful delivery of controller messages. If a faulty switch on the control path starts dropping controllers' messages, the next-in-line switches and dislocated controllers eventually start suspecting failed connections to the unreachable controllers. The switches eventually raise an alarm at the REASSIGNER. The REASSIGNER then accordingly marks the unreachable controllers as "unavailable". Only a limited number of "unavailable" controllers is tolerated by our design (governed by the $F_A$ parameter). Thus, both failures coming from unavailable controllers, as well as from the unavailable switches that forward the control packets to-and-from unreachable controllers are accounted by same parameter $F_A$.

\emph{ii) Re-adaptation to the optimal network configuration:} After the detection of a failed or malicious controller, the \emph{tamper-proof} entity REASSIGNER recomputes the controller-to-switch connection assignments in order to achieve the optimal network configuration given the current number of correct and incorrect controllers. The recomputed assignments are distributed to the controllers and switches in order to reduce the reconfiguration delay and messaging overhead in the control plane. The definition of the optimal configuration of a network and the reassignment logic is elaborated in Section \ref{ilp}.   

\subsection{System Elements in MORPH}
\label{implementedmodules}

\begin{figure}[htb]
	\centering
	\includegraphics[width=0.5\textwidth]{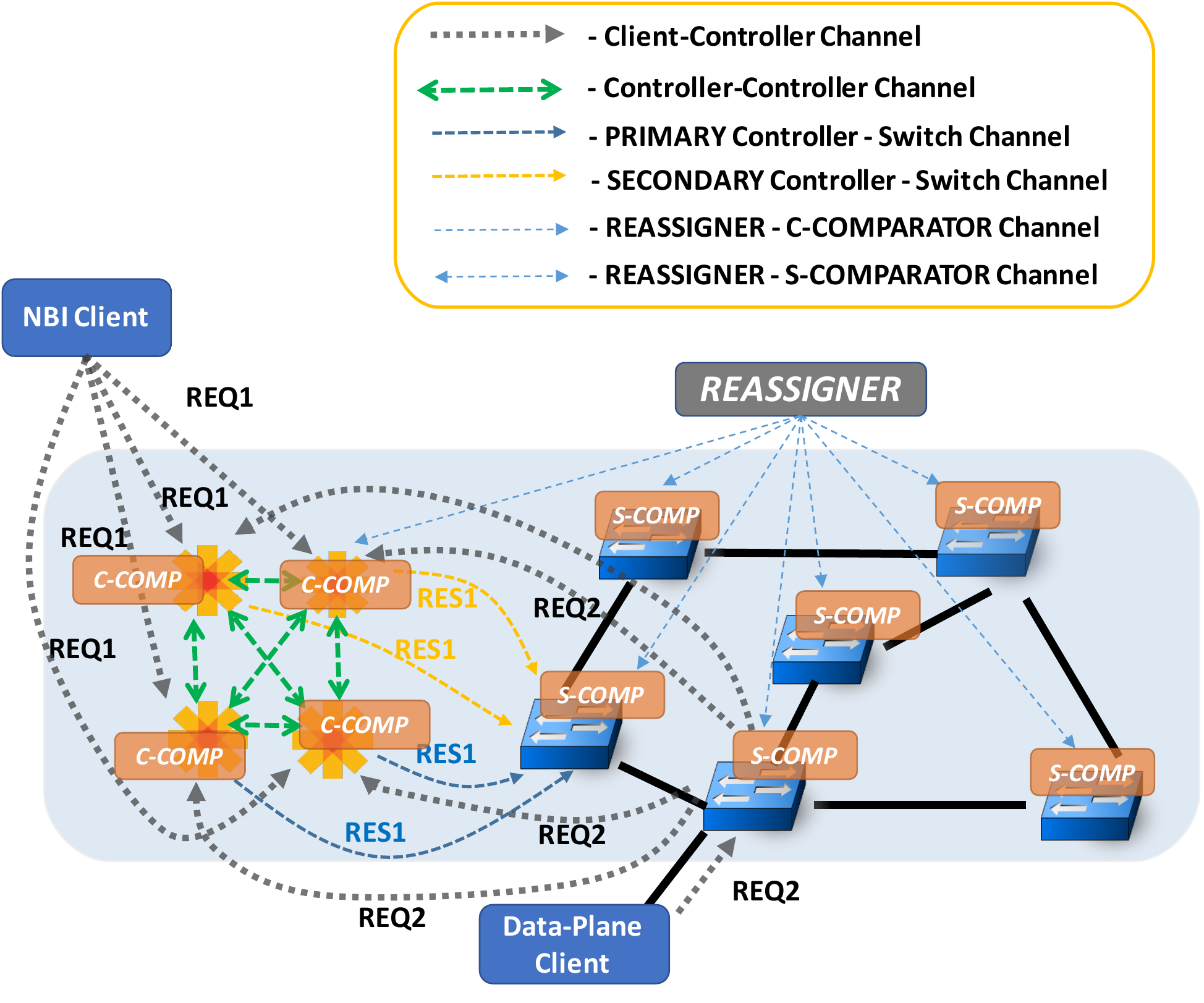}
	\caption{MORPH architecture comprising the: i) SDN controllers that host the C-COMPARATOR; ii) switches that host the S-COMPARATOR; iii) northbound (NBI) and data plane clients; and iv) REASSIGNER element. REASSIGNER is in charge of dynamically recomputing the controller-switch assignment after a reported controller failure discovery by S-COMPARATOR.}
	\label{fig:architecture}
\end{figure}

In order to establish and evaluate the mentioned design goals, the following elements are considered and introduced in the MORPH system architecture depicted in Fig. \ref{fig:architecture}.

\textbf{\emph{Northbound (NBI) and data plane clients}}: The SDN application clients requesting a particular controller service. NBI clients are aware of the controllers but not of their roles w.r.t. switch assignment. Thus, NBI clients report their requests to all available controllers. The data plane clients feed their requests to a well-defined controller group address. The data plane client requests are then handled by the neighboring edge switch and are encapsulated as a \emph{packet-in} message (e.g. using OpenFlow \cite{mckeown2008openflow}) and delivered only to the PRIMARY and SECONDARY controllers of that particular switch.

\textbf{\emph{S-COMPARATOR}}: A switch mechanism that collects and compares the configuration outputs generated at each active PRIMARY or SECONDARY controller instance associated with the switch hosting the C-COMPARATOR instance. On discovery of inconsistencies, the S-COMPARATOR reports the conflicting configurations to the \emph{REASSIGNER} entity. We assume a tamper-proof operation of the S-COMPARATOR (e.g., realized using the Intel SGX enclaves \cite{sgx}). S-COMPARATOR cannot be implemented with current OpenFlow logic, but requires one additional software agent, responsible for comparison of arriving control messages. This requirement is, however, not unique to MORPH and holds for any BFT-enabled system where switches take over the comparison logic, including \cite{mohan2017primary} and \cite{li2014byzantine}. Advances of programmable switching hardware and languages, such as P4 \cite{bosshart2014p4}, could enable for a more flexible implementations of S-COMPARATOR. A running instance of S-COMPARATOR is assumed in each active switch in MORPH deployment (refer to Fig. 2).

\textbf{\emph{C-COMPARATOR}}: A controller mechanism that collects and compares the controller configuration messages generated at each active controller instance. In the controller hosting SDA applications, it withholds the propagation of switch configuration messages as long as the majority of controller updates remains inconsistent. After collecting a majority of consistent messages, it updates the underlying controller's internal state configuration to reflect the configuration update (e.g., it updates the status of resource reservations). A running instance of C-COMPARATOR is assumed in each active controller in MORPH deployment (ref. Fig. 2).

\textbf{\emph{REASSIGNER}}: The REASSIGNER is in charge of detecting the incorrect controllers. It is furthermore able to find an optimal controller-to-switch assignment so to exclude the incorrect controllers and report the updated controller-switch list assignments at the controllers and switches. The REASSIGNER is triggered every time a controller is suspected by an S-COMPARATOR. The trigger is executed independent of the cause of failure (i.e. the discovery of a Byzantine or unavailable controller). However, the effect of discovery of the cause of failure considerably affects the selection of controllers considered in the assignment. The proposed controller-switch assignment solver is aware of the free capacities of the controllers as well as of the worst-case delays between the controllers and switches and in between the controllers, and is able to consider the related worst-case bounds when executing an optimal assignment. We detail the assignment procedure in Sec. \ref{ilp}.  We cannot trust a single REASSIGNER instance to be correct. To make the REASSIGNER resistant against malicious and availability failures, similarly to controllers, it must be implemented as a group of replicas that also carry out a BFT agreement protocol after each successful reassignment. For brevity, in the rest of the paper, we assume a tamper-proof operation of the REASSIGNER (e.g., realized using the Intel SGX enclaves \cite{sgx}) and focus our attention on solving the Byzantine faults in the context of SDN controllers only.

\subsection{Application Types}
\label{applicationtypes}
We distinguish two different types of SDN applications, \emph{state-independent (SIA)} and \emph{state-dependent applications (SDA)}. SIA refers to all SDN applications which process client's requests independent of the actual network state. Hence, given a repeated client input, a correct SIA application always generates a constant, semantically equal response. One representative of SIA is the \emph{hop-based shortest path routing} where, assuming no topology changes, the shortest path between the hosts remains the same, regardless of the current link and node utilization. Contrary to SIA, SDA refers to all applications which base their decision-making on the current network state. Hence, if we consider typical \emph{load balancing routing}, two identical requests (e.g. path requests) could generate completely different responses (i.e. different routes) from the controller, so to optimize for the total \emph{current} resource utilization given the instantaneous network state. 

\subsection{Necessity of the Controller State Synchronization}
\label{necessity}
\emph{Consensus across SDA instances:} In a resource-based SDA, such as a routing application, where the cost of each link depends on the current network state (e.g. link utilization), the problem might arise if two scattered clients (e.g. \emph{client A} and \emph{client B}) request the same path to distributed controllers at approximately the same time. Due to the propagation and processing delays, the controllers in the proximity of \emph{client A} and the controllers in the proximity of \emph{client B} could process the corresponding requests in a \emph{non-deterministic order}. Since the network state is updated after processing each request, even the correct controllers could produce inconsistent responses to the switches. MORPH solves this problem by \emph{delaying} the correct controllers' configuration responses to the switches until a sufficient number of controller messages generated by PRIMARY and SECONDARY set, required to reach a common \emph{consensus} for the client request, is collected. In order to achieve the consensus: i) all PRIMARY and SECONDARY controllers first compute a response immediately after receiving the client request; ii) they next store it in an internal database and; iii) exchange their decisions using the any-to-any C2C channels. Finally, the \emph{consensus} is reached for a controller's C-COMPARATOR component when it receives at least $\lfloor(Req_P+Req_S)/2\rfloor+1$ identical responses for a given request, where $Req_P$ is the current number of required PRIMARY controllers and $Req_S$ is the current number of required SECONDARY controller assignments per switch.

\emph{Proxying mechanism in SDA and SIA instances:} Furthermore, both in the SDA and SIA routing applications, the establishment of certain long paths is not achievable without the C2C state synchronization. For example, a client could initiate a routing request via its neighbouring edge switch $S_A$ to the set of $S_A$'s PRIMARY and SECONDARY controllers. However, if the determined path contains a switch which is not assigned to the same set of PRIMARY and SECONDARY controllers as with $S_A$, the configuration of the flow rules by the computing controller would not be allowed. Therefore, when an actual controller, which is assigned the role of PRIMARY or SECONDARY controller for the target switch, compares and validates the new configuration on the C2C channel, it decides to dispatch a response to the affected switches. The proxying mechanism could alternatively be omitted by enforcing the edge switch to forward the requests for SDN applications that operate globally on the data plane, to all available controllers. However, this design intuitively does not scale well with the number of controllers as the processing power of controllers and switches is limited \cite{kuzniar2015you}.

	\section{Enabling BFT Operation of a Distributed SDN Control Plane}
Next we detail the system flow and the algorithms behind MORPH. In the second part, we formulate the objectives and constraints for the Integer Linear Program (ILP) that dynamically reassigns the controller-switch connections during runtime, as part of the REASSIGNER logic.

\subsection{Algorithm}
\label{algorithm}

\begin{figure*}[htb]
	\centering
	\includegraphics[width=0.8\textwidth]{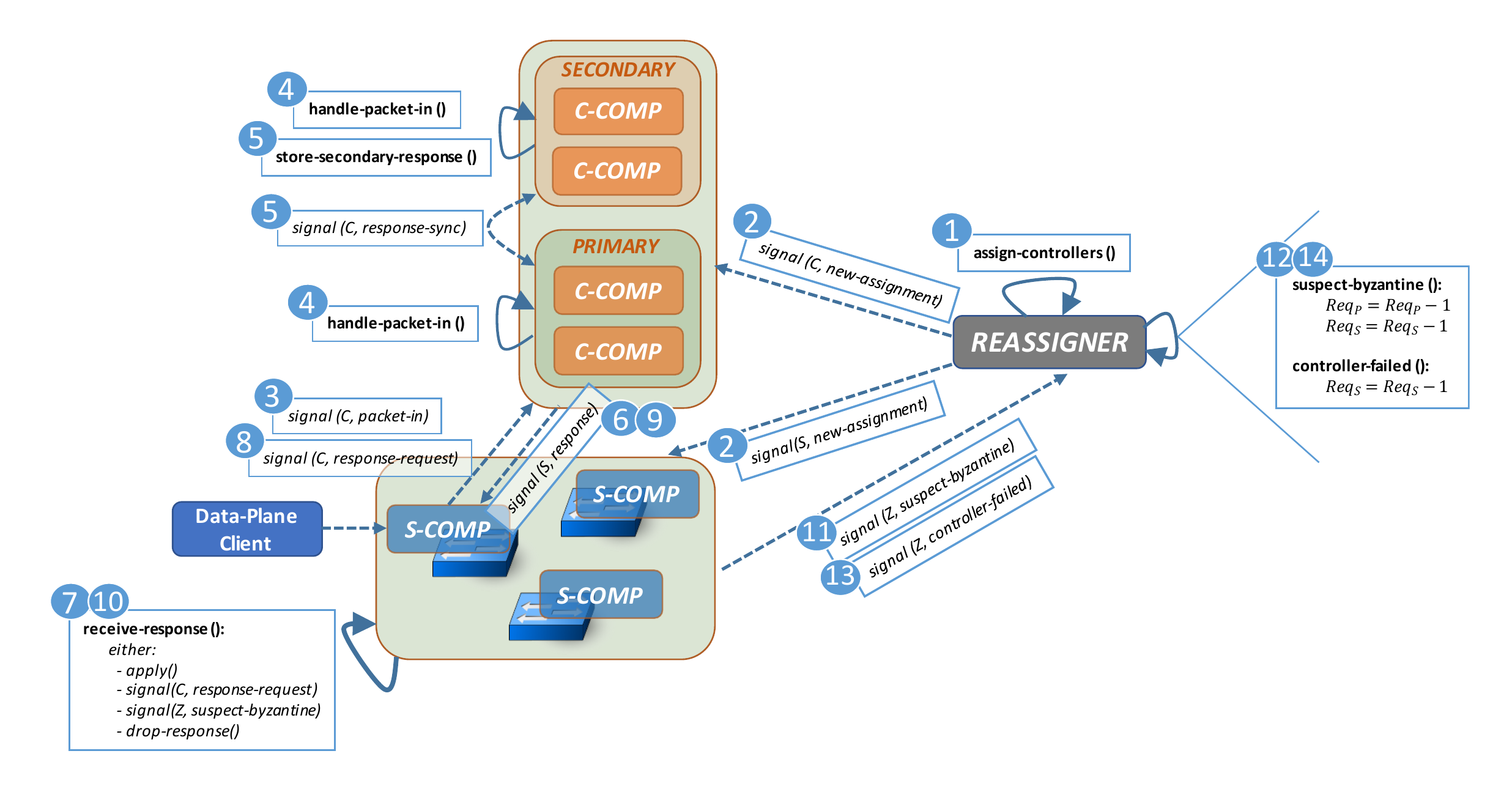}
	\caption{A simplified visual representation of the system workflow and the distributed algorithm execution as defined in Sec. \ref{algorithm}. The portrayed workflow depicts the steps of: (1-2) controller-switch assignment; (3-5) client request dissemination and handling in the PRIMARY and SECONDARY C-COMPARATOR instances; (6-7, 8-10) dissemination and handling of PRIMARY and SECONDARY controller responses in the affected S-COMPARATOR instances, respectively; (11-12, 13-14) dissemination and handling of Byzantine and/or availability-induced failures in the REASSIGNER, respectively.}
	\label{fig:algorithm}
\end{figure*}

The attached algorithms describe the operation of MORPH in more detail. A simplified visual representation and system workflow involving the introduced algorithms is provided in Fig. \ref{fig:algorithm}. We distinguish the following eight steps:
\begin{enumerate}[leftmargin=*]
		\item Given a list of controllers and switches, as well as the list of clients and their worst-case capacity requirements, REASSIGNER executes the initial assignment of control plane connections. It considers the unidirectional delay as well as the controller capacity constraint. As described in Sec. \ref{ilp}, the REASSIGNER minimizes the total number of active controllers when assigning switches to controllers. This process is embodied in Lines 1-2 of Alg. \ref{alg:reassigner-algorithm}.
	\item The controller-switch assignment lists are distributed to the SDN controllers and switches. Switches are thus assigned their PRIMARY and SECONDARY controllers. From now on, any received configuration message initiated by a remote controller is queued for the evaluation in the switch if and only if the configuration message was initiated by a controller that belongs to either the PRIMARY or SECONDARY controller set of that switch. Otherwise, the configuration message is rejected and dropped.
	\item An end-client sends off their request to the SDN controller, i.e. a request for a computation of a QoS-constrained path, a load-balancing request etc. Northbound clients send their requests directly to all controllers, while data plane clients stay unaware of the location of the controllers, so to simplify the client-side / user logic \cite{sardis2016can}. The data plane clients feed their requests directly into the network. The next-hop edge switch then intercepts and proxies the request to its assigned PRIMARY and SECONDARY controllers.
	\item The PRIMARY and SECONDARY controllers of the switch which initiated the client request compute the corresponding configuration response and decide to apply the computed response configuration in the affected switches. 	
		We enable the selection of a flexible trade-off between the switch configuration time overhead (response time) and the generated control plane load. In general, the controller instances assigned to a switch are either of the PRIMARY or the SECONDARY role. Depending on the point in time the SECONDARY controller instances decide to forward the computed configuration responses to their switches, the overhead in terms of response time and control plane load may vary. Therefore, we define the following two models: 
	\begin{itemize}
	\item \emph{NON-SELECTIVE}: Propagation of computed configurations from SECONDARY controllers to switches initiates immediately after receiving the client request. In case of the SDA, new configurations are sent to switches only after reaching the majority consensus on the new configuration message, as explained later in the text.
	\item \emph{SELECTIVE} model imposes an additional step of buffering the intermediate result (either locally computed in the case of SIA or the majority result in the case of SDA applications) and forwarding the result to the switch on-demand, whenever the switch requests additional configuration messages from its SECONDARY controllers.
	\end{itemize}

 In the case of the \emph{SELECTIVE} model, only the controllers which are the PRIMARY controllers of the to-be-configured switches forward their request directly to the switch. In the case of the \emph{NON-SELECTIVE} model, the SECONDARY controllers also proactively forward their configuration messages to the switches. This differentiation is depicted in Lines 9-16 of the SIA-specific Alg. \ref{controller-stateless-algorithm} and Lines 11-18 of the SDA-specific Alg. \ref{controller-stateful-algorithm}. 
	
		\emph{State-independent applications (SIA)}: In case of SIA, the controllers forward the computed configuration messages to the target switches if they are assigned either the PRIMARY or SECONDARY role for the target switch. If the controller that computes the configuration is assigned neither role for the target switch, the configuration result is forwarded to the switch only after a consensus is achieved on the actual PRIMARY/SECONDARY controllers of the switch. For SIA, consensus is achieved after collecting $Req_P$ identical messages on the PRIMARY/SECONDARY instances. $Req_P$ denotes the currently required number of assigned PRIMARY controllers per switch.
		
		\emph{State-dependent applications (SDA)}: In case of SDA, all correct controllers must first reach consensus on their common internal state update. To this end, they deduce the majority configuration message (in Lines 8-9 of Alg. \ref{controller-stateful-algorithm}). Majority configuration response is necessary in order to omit the possibility of false positive detection, where \emph{correct} controllers potentially become identified as faulty instances, following a temporary controller state de-synchronization during runtime, as discussed in Sec. \ref{soaissues}. After determining the majority response, the controllers send the configuration message to the switches (as in Lines 10-16 of Alg. \ref{controller-stateful-algorithm}). For SDA, the total number of required matching messages to apply a controller-state update is $\lfloor(Req_P+Req_S)/2\rfloor+1$, where $Req_S$ denotes the currently required number of assigned SECONDARY controllers per switch. The total number of exchanged messages is the same as in SIA, however, reaching a consensus before dispatching the configuration messages increases the overall reconfiguration time, as shown in Section VIII.

		The controllers that serve the client request synchronize their configuration messages with the PRIMARY and SECONDARY controllers of the affected switch. Thus, the PRIMARY and SECONDARY controllers of the affected switch apply new switch configurations only after a sufficient number of matching messages were generated at the remote controller cluster members that have executed the computation. Alternatively, the switches may distribute the client requests to ALL SDN controllers and thus omit the round trip resulting from the corner-case described above. This incurs an additional message load for any case that is not the corner-case situation described above.

	\item The S-COMPARATORS collect the configuration requests and decide after $Req_P$ consistent (equal) PRIMARY messages to apply the configuration locally. In the SELECTIVE scenario where inconsistent messages among the $Req_P$ PRIMARY messages are detected, the switches contact the SECONDARY replicas to fetch additional $Req_S$ responses - as depicted in Line 10 of Alg. \ref{switch-algorithm}. After receiving $Req_P$ consistent (equal) messages (Line 12-15 of Alg. \ref{switch-algorithm}), the switches apply the new configuration (Line 15). Thus, the overhead of collection of $Req_P$  consistent messages dictates the worst-case for applying new switch configurations.

	\item 	After discovery of an inconsistency among the controller responses (Lines 9-10 of Alg. \ref{switch-algorithm}) or a failure of an assigned controller (Line 24-25), the S-COMPARATORS wait until $Req_P+Req_S$ messages are collected, before addressing the REASSIGNER with the controller IDs and the conflicting messages (Line 20 of Alg. \ref{switch-algorithm}). In the case when a controller instance has failed, the duration of the time the switch waits before contacting the REASSIGNER corresponds to the worst-case failure detection period (dictated by the underlying failure detector, e.g. the $\phi$-Accrual \cite{phi-accrual} detector). If a switch suspects that a controller assigned to it has failed, it notifies the REASSIGNER as per Line 25 of Alg. \ref{switch-algorithm} and Line 13 of Alg. \ref{alg:reassigner-algorithm}.
	\item The REASSIGNER compares the inputs and deduces the majority of correct responses received for the conflicting client request. If a Byzantine failure of a controller is suspected, both $Req_P$ and $Req_S$ are decremented by one for each malicious controller. If a controller is marked as unavailable (independent of the source of failure and correctness of the controller), only the required number of SECONDARY controllers per-switch $Req_S$ is decremented by one. Lines 4-17 of Alg. \ref{alg:reassigner-algorithm} contain this differentiation.
\item Based on the updated $Req_P$ and $Req_S$ controllers deduced during runtime, the REASSIGNER computes the new optimal assignment and configures the switches and controllers with the new controller-switch assignment lists. Steps 3-7 repeat until all $F_M$ malicious and $F_A$ unavailable controllers are eventually identified.
\end{enumerate}

By lowering the number of maximum required PRIMARY and SECONDARY controller assignments per switch, the REASSIGNER minimizes the total control plane overhead in terms of the packet exchange in both controller-to-controller and controller-to-switch channels, as well as the time required to confirm new controller and switch state configurations. 

\begin{algorithm}[htb]
	\caption{REASSIGNER: Controller-switch assignment}
	\label{alg:reassigner-algorithm}
	\hspace*{\algorithmicindent} \textbf{Notation}: \\
	\hspace*{\algorithmicindent} $\mathcal{S}$ Set of available SDN switches\\
	\hspace*{\algorithmicindent} $\mathcal{C}$ Set of available SDN controllers\\
	\hspace*{\algorithmicindent} $\mathcal{B}$ Set of detected blacklisted SDN controllers\\
	\hspace*{\algorithmicindent} $\mathcal{A}_p$ Set of PRIMARY controller-switch assignments\\
	\hspace*{\algorithmicindent} $\mathcal{A}_s$ Set of SECONDARY controller-switch assignments\\
	\hspace*{\algorithmicindent} $Req_{P}$ No. of required PRIMARY controllers per switch\\
	\hspace*{\algorithmicindent} $Req_{S}$ No. of req. SECONDARY controllers per switch\\
	\hspace*{\algorithmicindent} $F_M$ Maximum number of tolerated malicious controller failures \\
	\hspace*{\algorithmicindent} $F_A$ Maximum number of tolerated unavailability controller failures \\

	\hspace*{\algorithmicindent} \textbf{Initial variables}: \\
	\hspace*{\algorithmicindent} $Req_{P} = F_M+1$\\
	\hspace*{\algorithmicindent} $Req_{S} = F_A+F_M$\\ 
	\label{shuffle}
	\begin{algorithmic}[1]
		\Procedure{Controller-Switch Assignment}{}
		\State $(\mathcal{A}_p, \mathcal{A}_s) :=$ \emph{Assign-Controllers} $(\mathcal{S}, \mathcal{C})$ 
		\\
		\BState \textbf{ upon event} \emph{suspect-byzantine} $<\mathcal{C}_{m}>$ \textbf{do}
		\State $\mathcal{C} \leftarrow \mathcal{C} \setminus \mathcal{C}_{m}$
		\State $\mathcal{B} \leftarrow \mathcal{B} \cup \mathcal{C}_{m}$
		\State $Req_{P} = Req_{P}-1$
		\State $Req_{S} = Req_{S}-1$
		\State $(\mathcal{A}_p, \mathcal{A}_s) :=$ \emph{Assign-Controllers} $(\mathcal{S}, \mathcal{C}, Req_{P}, Req_{S})$ 
		\State \emph{signal} ($S$, \emph{new-assignment}$<\mathcal{A}_p, \mathcal{A}_s>)$
		\State \emph{signal} ($C$, \emph{new-assignment}$<\mathcal{A}_p, \mathcal{A}_s, \mathcal{B}>)$
		\\
		\BState \textbf{ upon event} \emph{controller-failed} $<\mathcal{C}_{f}>$ \textbf{do}
		\State $\mathcal{C} \leftarrow \mathcal{C} \setminus \mathcal{C}_{f}$
		\State $\mathcal{B} \leftarrow \mathcal{B} \cup \mathcal{C}_{m}$
		\State $Req_{S} = Req_{S}-1$
		\State $(\mathcal{A}_p, \mathcal{A}_s) :=$ \emph{Assign-Controllers} $(\mathcal{S}, \mathcal{C}, Req_{P}, Req_{S})$ 
		\State \emph{signal} ($S$, \emph{new-assignment}$<\mathcal{A}_p, \mathcal{A}_s>)$
		\State \emph{signal} ($C$, \emph{new-assignment}$<\mathcal{A}_p, \mathcal{A}_s, \mathcal{B}>)$
		\EndProcedure
	\end{algorithmic}
\end{algorithm}

\begin{algorithm}[htb]
	\caption{C-COMPARATOR: Handling of state-independent application (SIA) requests in the SDN controller}
	\label{controller-stateless-algorithm}
	\hspace*{\algorithmicindent} \textbf{Notation}: \\
	\hspace*{\algorithmicindent} $P$ Client request (e.g. flow request) initiated at switch $S_C$\\
	\hspace*{\algorithmicindent} $\mathcal{C}$ The set of available SDN controllers\\
	\hspace*{\algorithmicindent} $C_i$ The local controller instance\\
	\hspace*{\algorithmicindent} $C_r$ The remote controller instance\\
	\hspace*{\algorithmicindent} $R_{P, S_i}$ Configuration response intended for switch $S_i$\\
	\hspace*{\algorithmicindent} $\mathcal{A}_p$ Set of primary controller-switch assignments\\
	\hspace*{\algorithmicindent} $\mathcal{A}_s$ Set of secondary controller-switch assignments\\
	\hspace*{\algorithmicindent} $A_{S_i, C_i}$ Connection assignment variable for controller-switch pair $(S_i, C_i)$ \\
	\label{controller}
	\begin{algorithmic}[1]
		\Procedure{Handle client request}{}
		\BState \textbf{ upon event} \emph{packet-in} $<S_C, P>$ \textbf{do} 
		\State $R_{P}:=$ \emph{handle-packet-in} $(P)$
		\State \emph{signal} ($\mathcal{C}$, \emph{response-sync}$<C_i, R_{P}>$)
		\\	
		\BState \textbf{ upon event} \emph{ response-sync} $<C_r, R_{P}>$ \textbf{do}
		\If {\emph{is-sia-consensus-reached}$(\mathcal{R}^P)$ \textbf{or} $C_i == C_r$}
			\ForAll {$S_i \in$ \emph{affected-switches($R_{P}$)}}
					\If {mode == \emph{non-selective}}
						\If {$A_{S_i, C_i} \in \mathcal{A}_p \cup \mathcal{A}_s$}
							\State \emph{signal} $(S_i, response<C_i, R_{P, S_i}>)$
						\EndIf
					\ElsIf {mode == \emph{selective}}
						\If {$A_{S_i, C_i} \in \mathcal{A}_p$}
							\State \emph{signal} $(S_i, response<C_i, R_{P, S_i}>)$
						\ElsIf {$A_{S_i, C_i} \in \mathcal{A}_s$}
							\State \emph{store-secondary-response}$(R_{P, S_i})$
						\EndIf
					\EndIf
			\EndFor
		\EndIf
		\\	
		\BState \textbf{ upon event} \emph{ response-request} $<S_i, P>$ \textbf{do} 
			\State \emph{signal} $(S_i, response<C_i, R_{P, S_i}>) $
		\EndProcedure
	\end{algorithmic}
\end{algorithm}

\begin{algorithm}[htb]
	\caption{C-COMPARATOR: Handling of state-dependent application (SDA) requests in the SDN controller}
	\label{controller-stateful-algorithm}
	\hspace*{\algorithmicindent} \textbf{Notation}: \\
	\hspace*{\algorithmicindent} $P$ Client request (e.g. flow request) initiated at switch $S_C$\\
	\hspace*{\algorithmicindent} $\mathcal{C}$ The set of available SDN controllers\\
	\hspace*{\algorithmicindent} $C_i$ The local controller instance\\
	\hspace*{\algorithmicindent} $C_r$ The remote controller instance\\
	\hspace*{\algorithmicindent} $R_{P, S_i}^{maj}$ Majority configuration for switch $S_i$\\
	\hspace*{\algorithmicindent} $\mathcal{R}^P$ Buffer containing controller responses for request $P$ \\
	\hspace*{\algorithmicindent} $\mathcal{A}_p$ Set of primary controller-switch assignments\\
	\hspace*{\algorithmicindent} $\mathcal{A}_s$ Set of secondary controller-switch assignments\\
	\hspace*{\algorithmicindent} $A_{S_i, C_i}$ Connection assignment variable for controller-switch pair $(S_i, C_i)$ \\
	\label{controller}
	\begin{algorithmic}[1]
		\Procedure{Handle client request}{}
		\BState \textbf{ upon event} \emph{packet-in} $<S_C, P>$ \textbf{do} 
		\State $R_{P}:=$ \emph{handle-packet-in} $(P)$
		\State \emph{signal} ($\mathcal{C}$, \emph{response-sync}$<C_i, R_{P}>$)
		\\	
		\BState \textbf{ upon event} \emph{ response-sync} $<C_r, R_{P}>$ \textbf{do} 
		\State $\mathcal{R}^P \leftarrow \mathcal{R}^P \cup <C_r, R_P>$
		\If {\emph{is-sda-consensus-reached}$(\mathcal{R}^P)$}
		\State $R_{P}^{maj} := $\emph{majority-response($\mathcal{R}^P$)}
			\ForAll {$S_i \in$ \emph{affected-switches($R_{P}^{maj}$)}}
				\If {mode == \emph{non-selective}}
					\If {$A_{S_i, C_i} \in \mathcal{A}_p \cup \mathcal{A}_s$}
						\State \emph{signal} $(S_i, response<C_i, R_{P, S_i}^{maj}>)$
					\EndIf
				\ElsIf {mode == \emph{selective}}
					\If {$A_{S_i, C_i} \in \mathcal{A}_p$}
						\State \emph{signal} $(S_i, response<C_i, R_{P, S_i}^{maj}>)$
					\ElsIf {$A_{S_i, C_i} \in \mathcal{A}_s$}
						\State \emph{store-secondary-response}$(R_{P, S_i}^{maj})$
					\EndIf
				\EndIf
			\EndFor
		\EndIf
		\\	
		\BState \textbf{ upon event} \emph{ response-request} $<S_i, P>$ \textbf{do} 
			\State \emph{signal} $(S_i, response<C_i, R_{P, S_i}^{maj}>) $
		\EndProcedure
	\end{algorithmic}
\end{algorithm}

\begin{algorithm}[htb]
	\caption{S-COMPARATOR: Processing of controller configuration messages in the switch}
	\label{switch-algorithm}
	\hspace*{\algorithmicindent} \textbf{Notation}: \\
	\hspace*{\algorithmicindent} $Req_{P}$ No. of required PRIMARY controllers per switch\\
	\hspace*{\algorithmicindent} $Req_{S}$ No. of req. SECONDARY controllers per switch\\
	\hspace*{\algorithmicindent} $\mathcal{R}^P$ The set of received responses for client request $P$ \\
	\hspace*{\algorithmicindent} $\mathcal{A}_p^{S_i}$ Set of PRIMARY controllers assigned to switch $S_i$ \\
	\hspace*{\algorithmicindent} $\mathcal{A}_s^{S_i}$ Set of SECONDARY controllers assigned to $S_i$ \\
	\hspace*{\algorithmicindent} $\mathcal{Z}$ The set of designated shufflers assigned to switch $S_i$ \\

	\begin{algorithmic}[1]
		\Procedure{Compare and apply controller configurations}{}
		\BState \textbf{ upon event} \emph{receive-response}$<C_j, R^P>$ \textbf{do}
		\State $\mathcal{R}^P \leftarrow \mathcal{R}^P \cup <C_j, R^P>$

		\If {$C_j \in \mathcal{A}_p^{S_i}$}
			\If {$|\mathcal{R}^P| == Req_{P}$}
				\State $\mathcal{R}_{majority} :=$ \emph{majority-responses($\mathcal{R}^P$)}
				\If {$|\mathcal{R}_{majority}| \geq Req_{P}$}
					\State \emph{apply} $(\mathcal{R}_{majority})$
				\Else
					\State{\emph{signal}$(\mathcal{A}_s^{S_i},$ \emph{response-request} $<S_i, P>)$}
				\EndIf
			\EndIf
		\ElsIf {$C_j \in \mathcal{A}_p^{S_i} \cup {A}_s^{S_i}$}
		\If {$Req_{P} < |\mathcal{R}^P| \leq Req_{P} + Req_{S}$}
		\State $\mathcal{R}_{majority} :=$ \emph{majority-responses}($\mathcal{R}^P$)
		\If {$|\mathcal{R}_{majority}| \geq Req_{P}$}
		\State \emph{apply} $(\mathcal{R}_{majority})$
		\ForAll{$<C_i, R_i> \in \mathcal{R}^P$}
		\If {$R_i \notin \mathcal{R}_{majority}$}
		\State $\mathcal{C}_{incnst} \leftarrow \mathcal{C}_{incnst} \cup C_i$
		\EndIf
		\EndFor
		\If{$C_i \neq \emptyset$}
		\State \emph{signal} $(\mathcal{Z},$ \emph{suspect-byzant} $<\mathcal{C}_{incnst}>$)
		\EndIf
		\EndIf
		\EndIf
		\Else
		\State \emph{drop-response}$(R^P)$
		\EndIf
		\\		
		\BState \textbf{ upon event} \emph{controller-failed}$<C_j>$ \textbf{do}
			\State \emph{signal} $(\mathcal{Z},$ \emph{controller-failed} $<\mathcal{C}_j>$)
		\EndProcedure
	\end{algorithmic}
\end{algorithm}

\subsection{REASSIGNER Logic (Integer Linear Program)}
\label{ilp}

REASSIGNER assigns the controller roles to switches on a per-switch basis. It takes a list of controllers and switches, their unidirectional delay and delay requirements, as well as the list of clients and the client request arrival rates as input. 

For brevity, we henceforth define a single objective for the assignment problem. Namely, the REASSIGNER minimizes the number of active controllers, so to lower the total overhead of controller-to-controller communication, while taking into consideration the maximum delay bound and available controller capacities when assigning the controllers to switches. The mentioned objective additionaly allows for implicit load-balancing of individual controller-switch connections, as long as all controllers are instantiated with an equal initial capacity.

Let $A_{C_i,S_j}^{P}$ and $A_{C_i,S_j}^{S}$ denote the active assignment of a controller instance $C_i$ to the switch $S_j$ as a PRIMARY/SECONDARY controller, respectively. Then $U_{C_i}$ denotes the active participation of the controller $C_i$ in the system:

\begin{equation}
	U_{C_i} = \begin{cases}
		1 & \mathrm{when} \sum\limits_{S_j\in\mathcal{S}}{A^P_{C_i,S_j}} + \sum\limits_{S_j\in\mathcal{S}}{A^S_{C_i,S_j}} > 0, \forall C_i \in \mathcal{C}\\
		0 & \mathrm{otherwise}
\end{cases}
\end{equation}

The objective function can then be formalized as:

\begin{equation}
	\label{objective-function}
	min \sum_{C_i\in\mathcal{C}}{U_{C_i}}
\end{equation}

In the following, we define the constraints of the corresponding ILP.

\emph{Minimum Assignment Constraint}: 
In order to tolerate $F_M$ Byzantine and $F_A$ unavailability failures, initially at time $t=0$ the REASSIGNER assigns $Req_P(0) = F_M+1$ PRIMARY controllers, and $Req_S(0) = F_M+F_A$ SECONDARY controllers per each switch. The values $Req_P$ and $Req_S$ are time-variant and are adapted on every discovered Byzantine/availability failure or a successful controller recovery:

\begin{equation}
	\begin{split}
		\sum\limits_{C_i\in\mathcal{C}}{A^P_{C_i,S_j}} \geq Req_P(t), \forall S_j \in \mathcal{S} \\
		\sum\limits_{C_i\in\mathcal{C}}{A^S_{C_i,S_j}} \geq Req_S(t), \forall S_j \in \mathcal{S} 
\end{split}
	\label{ilp-constraint-1}
\end{equation}

\emph{Unique Assignment Constraint}: Each controller $C_i$ may either be assigned the role of a PRIMARY \emph{or} SECONDARY, at maximum \emph{once} per switch $S_j$:
\begin{equation}
	A^P_{C_i,S_j} + A^S_{C_i,S_j} \leq 1, \forall C_i \in \mathcal{C}, S_j \in \mathcal{S}
	\label{ilp-constraint-2}
\end{equation}

\emph{Controller Capacity Constraint}: Let $P_{C_i}$ denote the total available controller's $C_i$ capacity. Let $L_{CL_k}$ and $L_{S_j}$ denote the load stemming from the northbound client $CL_k$ and the data plane edge switch $S_j$, respectively. The sum of the loads generated by the assigned switches at controller $C_i$ may not exceed the \emph{difference} of the \emph{total available controller's capacity} and the \emph{sum of the constant loads} stemming from the northbound clients (ref. Sec. \ref{systemmodel}):
\begin{equation}
	\begin{split}
	\sum\limits_{S_j\in\mathcal{S}}{A^P_{C_i,S_j}*L_{S_j}} + \sum\limits_{S_j\in\mathcal{S}}{A^S_{C_i,S_j}*L_{S_j}} \leq \\ P_{C_i} - \sum\limits_{CL_k\in\mathcal{CL}}{L_{CL_k}}, \forall C_i \in \mathcal{C} 
\end{split}
\label{ilp-constraint-3}
\end{equation}

\emph{Delay Bound Constraint}: 
We assume well-defined worst-case upper bounds for the unidirectional delays in controller-to-switch and controller-to-controller communication. Let $d_{C_i, S_j}$ and $d_{C_i, C_j}$ denote the guaranteed worst-case experienced delay for the controller-switch pair ($C_i$, $S_j$) and controller-controller pair ($C_i$, $C_j$), respectively. Given a maximum tolerable global upper bound delays $D_{C,S}$ and $D_{C,C}$, we define the related constraints as:
\begin{equation}
\begin{split}
	A^P_{C_i,S_j}*d_{C_i, S_j} \leq D_{C,S}, \forall C_i \in \mathcal{C}, S_j\in\mathcal{S}\\
	A^S_{C_i,S_j}*d_{C_i, S_j} \leq D_{C,S}, \forall C_i \in \mathcal{C}, S_j\in\mathcal{S}\\
	d_{C_i, C_j} \leq D_{C,C}, \forall C_i \in \mathcal{C} \setminus C_j, C_j\in\mathcal{C} \setminus C_i\\
\end{split}
\label{ilp-constraint-4}
\end{equation}

\subsection{Communication and Computation Complexity Analysis}
\label{subsec:comm_analysis}

\begin{table}[htb]
\centering
\caption{Notation used in specification of the communication overhead \label{table:communicationnotation}} 
\resizebox{\columnwidth}{!}{
\begin{tabular}{>{\arraybackslash}m{0.7cm} >{\arraybackslash}m{9cm}} 
\hline
Notation & Property \\  
\hline
CLI2S & Communication channels between clients and switches \\
S2C & Upstream communication channels between switches and controllers \\
C2C & Controller to controller communication channels \\
C2S & Downstream communication channels between controllers and switches \\
C2S & Downstream communication channels between controllers and switches \\
$m$ & Total number of data plane clients in the network\\
$\mu_{cli}$ & The average request rate of data plane clients\\
$|S_{cli}|$ & The average number of affected switches by data plane client's requests\\
$|C|$ & The total number of controllers, $|C| \geq 2F_M + F_A + 1$\\
$f_{M,A}$ & the current number of failed controllers\\
$E$ & Equal to 1/0 if MORPH is in NON-SELECTIVE/SELECTIVE mode\\
\hline
\end{tabular}
}
\end{table}

	Henceforth we analyze the communication overhead imposed by MORPH framework when handling data plane clients only (NBI clients are excluded for brevity). Multiple communication channels are present in the system:

\begin{itemize}
	\item \textit{CLI2S:} Every request generated by a data plane client is first forwarded to the corresponding edge switch, therefore, if the average request generation rate of $m$ data plane clients is $\mu_{cli}$, then $m \mu_{cli}$ is the total request generation rate on \textit{client} to \textit{switch} channels (i.e., CLI2S).

	\item \textit{S2C:} Upon receiving the request, the edge switch forwards it to its PRIMARY and SECONDARY controllers, hence, the total message rate on the \textit{upstream} communication channels between \textit{switches} and \textit{controllers} is $(Req_P + Req_S) m \mu_{cli}$. In the best-case, all faulty controllers were detected, hence the total number of PRIMARY and SECONDARY controllers is reduced to $Req_P=1$ and $Req_S=0$, while in the worst-case none of the faulty controllers failed yet, i.e., the inital values still hold $Req_P = F_M + 1$ and $Req_S = F_M + F_A$. 

	\item \textit{C2C:} For every request received from the switches, its PRIMARY and SECONDARY controllers calculate the corresponding response (i.e., a forwarding path) and forward it to all of the \textit{other available} (non-faulty) controllers in the network (i.e., $|C| -f_{M,A} - 1$). The total dynamic rate rate on the \textit{controller} to \textit{controller} channel is $(|C| -f_{M,A} - 1) (Req_P + Req_S) m \mu_{cli}$. In the best-case, all failed controllers were detected, thus $f_{M,A} = F_M + F_A$, however, in the worst-case, none of the controllers failed, i.e., $f_{M,A} = 0$.

	\item \textit{C2S:} After the consensus is reached for the corresponding request, in the case of NON-SELECTIVE mode all PRIMARY and SECONDARY controllers issue the reconfiguration responses to all affected switches (i.e., variable is $E=1$), while in the SELECTIVE mode, only PRIMARY controllers issue the responses (i.e., $E=0$). If we consider that on average $|S_{cli}|$ switches are reconfigured based on the clients requests, the total number of responses issued by \textit{controllers} towards \textit{switches} is $|S_{cli}|  (Req_P + E Req_S) m \mu_{cli}$. In the best-case, all faulty controllers were detected (i.e., $Req_P = 1$) and the working mode is SELECTIVE $E = 0$, thus only $|S_{cli}| m \mu_{cli}$ messages are needed. While in the worst-case, the initial case, $Req_P = F_M + 1$, $E=1$, $Req_S = F_M + F_A + 1$.
\end{itemize}

\begin{table*}[htb]
	\caption{Exchanged number of messages on each communication channel} \label{table:communication1} 
\begin{center}
\scalebox{0.8}{%
\begin{tabular}{|l|l|l|l|l|l|} 
\hline
\textbf{Channel} & \textbf{MORPH} & \textbf{Best-Case MORPH} & \textbf{Worst-Case MORPH} & \textbf{No Fault-Tolerance} & \textbf{\cite{mohan2017primary}} \\  
\hline\hline
CLI2S & $m \mu_{cli}$ & $m \mu_{cli}$ & $m \mu_{cli}$ & $m \mu_{cli}$ & $m \mu_{cli}$ \\
\hline
S2C & $(Req_P + Req_S) m \mu_{cli}$ & $m \mu_{cli}$ & $(2F_M + F_A + 1) m \mu_{cli}$ & $m \mu_{cli}$ & $(2(F_M + F_A) + 1) m \mu_{cli}$ \\
\hline
C2C & $(|C|-f_{M,A}-1)(Req_P + Req_S)m \mu_{cli}$ & $(|C| - F_M - F_A - 1) m \mu_{cli}$ & $(|C| - 1) (2F_M + F_A + 1) m \mu_{cli}$ & - & - \\
\hline
C2S & $|S_{cli}|(Req_P + E Req_S) m \mu_{cli}$ & $|S_{cli}| m \mu_{cli}$ & $|S_{cli}| (2F_M + F_A + 1) m \mu_{cli}$ & $|S_{cli}| m \mu_{cli}$ & $|S_{cli}| (2(F_M + F_A) + 1) m \mu_{cli}$ \\
\hline
\end{tabular}
}
\end{center}
\end{table*}

The summary of the communication overhead is presented in Table \ref{table:communication1}, with corresponding notation summarized in \ref{table:communicationnotation}. Table \ref{table:communication1} also presents the required number of messages in case when there is \textit{no fault-tolerance}, i.e., in this case we assume that the network is handled by a single SDN controller. Furthermore, we also present the communication overhead specific to the design presented in a related work~\cite{mohan2017primary}. It can be observed that MORPH requires the same or lower amount of messages on CLI2S, S2C, C2S channels depending on the current network state. Furthermore, as mentioned in the Sec. IV, controller-to-controller communication was not discussed in~\cite{mohan2017primary}, thus SDA are not correctly handled as in the case of MORPH.

\section{Evaluation Methodology}
\label{eval}
In this section we present our evaluation methodology including the target SDN application, the evaluated topologies and the overall scenario. Furthermore, we elaborate our emulation approach in more detail.

\subsection{Application model}
For our SDN application we have implemented a centralized resource-based path finding application, based on Dijkstra's algorithm \cite{dijkstra1959note}. We deploy the routing application once per each controller instance, so to cater to the SPOF issue.

In the case of SIA realization, the weight of each link is set to a constant link delay value corresponding to the propagation delay. Hence, the application is stateless and computes the same shortest path between two arbitrary hosts as long as no topology changes occur. Thus, as soon as our SDN controller application receives a client request, it computes the new path, signals the path configuration (i.e. OpenFlow \emph{FlowMod} updates) to the affected switches, and notifies the other controllers for state synchronization purposes. 

For the SDA case, we consider load-balanced routing where the cost of each link depends on the current link load. A correct distribution of switch configuration thus necessitates a consensus across the set of correct SDN controllers, so in order to omit the overloading of particular links (ref. Sec. \ref{necessity}). 

\subsection{Scenarios}

\begin{figure}[htb]
	\centering
	\subfloat[Internet2 topology \cite{hock2014poco}]{
		\centering
		\includegraphics[width=0.35\textwidth]{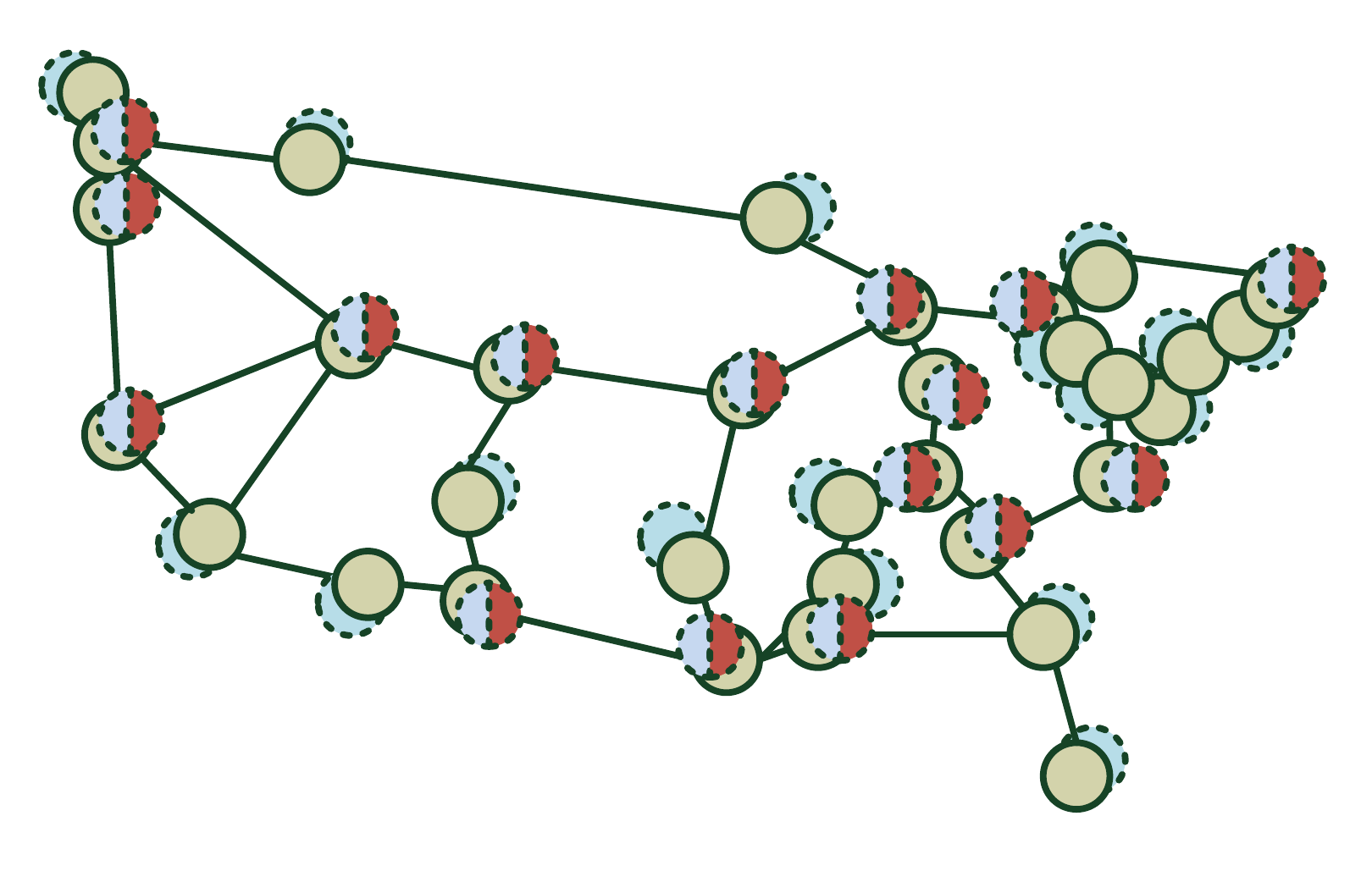}
		\label{fig:topo1}
	}
	\newline
	\subfloat[Fat-tree topology \cite{huang2017dynamic}]{
		\centering
		\includegraphics[width=0.35\textwidth]{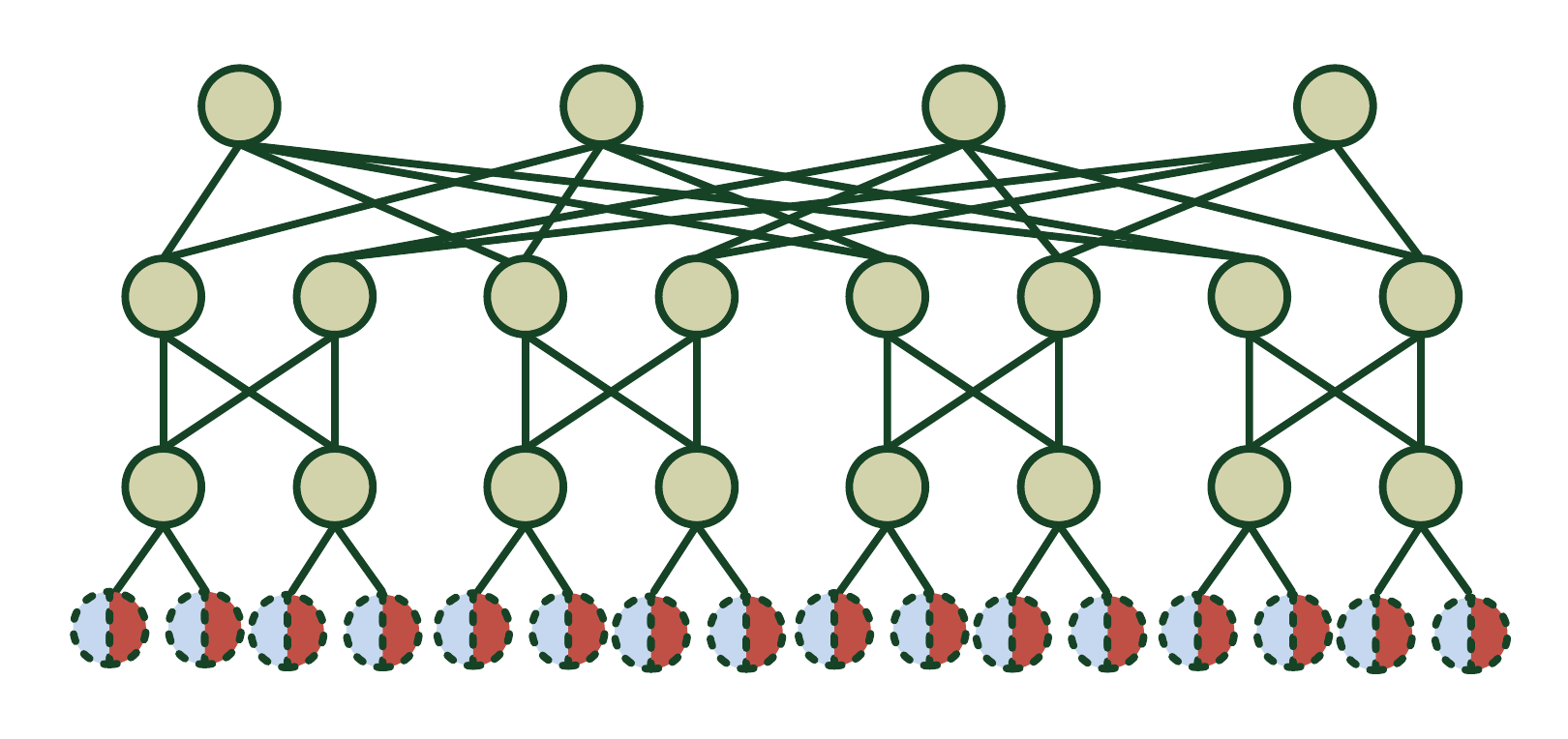}
		\label{fig:topo2}
	}
	\caption{Exemplary network topologies and controller placements used in the evaluation of MORPH. Elements highlighted in green and blue represent the switches and clients, respectively. Red elements are the controller instances placed as per \cite{hock2014poco, huang2017dynamic}. Each switch instance in the Internet2 topology is allocated a client, while the fat-tree topology hosts clients at the leaf switches.}
	\label{topologies}
\end{figure}

To validate our claims in a realistic environment, we have emulated the Internet2 Network Infrastructure Topology\footnote{Internet2 Advanced Networking - \url{https://www.internet2.edu/products-services/advanced-networking/}}, as well as a standard fat-tree data-center topology. Both topologies are depicted in Fig. \ref{fig:topo1} and Fig. \ref{fig:topo2}, respectively. The evaluated Internet2 and fat-tree topologies encompass 34 and 20 switches, respectively. We have enabled a configuration of MORPH to support up to $F_M=5$ and $F_A=5$ availability-related controller failures. Thus, to allow for a Byzantine fault-tolerant operation, $2F_M+F_A+1=16$ controller processes were deployed. We varied $F_M$ and $F_A$ individually from $1$ to $5$, so to allow for an evaluation of the overhead of added robustness against either type of failure. The overhead of our design scales with the arbitrary number of tolerated failures $F_M$ and $F_A$ and is independent of the number of deployed switches. Each SDN controller implements the logic to execute the routing task as well as the C-COMPARATOR component that compares the controller-to-controller synchronization inputs and notifies the switches of new path configurations. An S-COMPARATOR agent is hosted inside each switch instance, and is enabled to listen for remote connections. All communication channels (ref. Fig. \ref{fig:architecture}) are realized using TCP with enabled Nagle's algorithm \cite{minshall2000application}.

A realistic, in-band control plane channel was realized at all times. To realize the switching plane, a number of interconnected Open vSwitch\footnote{Open vSwitch - \url{https://www.openvswitch.org/}} v2.8.2 virtual switches were instantiated and isolated in individual Docker containers. To reflect the delays incurred by the length of the optical links in the geographically scattered Internet2 topology, we assume a travel speed of light of \begin{math} 2\cdot106 km/s \end{math} in the optical fiber links. We then derive the link distances from the publicly available geographical Internet2 data\footnote{Internet2 topological data (provided by POCO project) - \url{https://github.com/lsinfo3/poco/tree/master/topologies}} and inject the propagation delays using Linux's Traffic Control (\emph{tc}) tool. The links of the fat-tree topology only posses the inherit processing and queuing delays. The arrival rates of the incoming service embedding requests were modeled using a negative exponential distribution \cite{huang2017dynamic}. In the fat-tree topology, each leaf-switch was connected to $2$ client instances, bringing the total number of clients up to $16$. The Internet2 topology deployed one client per switch.

\emph{Controller placement}: In the Internet2 topology, we leverage a controller placement that allows for a high robustness against the controller failures. We consider the exemplary placements introduced in \cite{hock2014poco}. Since the accompanying controller placement framework\footnote{Pareto Optimal Controller Placement (POCO) GitHub - \url{https://github.com/lsinfo3/poco}} was unable to solve the placement problem for a very high number of controllers (i.e. up to $16$ in our case), we have executed the problem by placing up to $5$ controllers multiple times for the same topology, targeting different optimization objectives (incl. response time, maximized coverage in the case of failures, load balancing etc.). We then ranked the unique nodes based on their placement preference and have selected the highest-ranked 16 nodes to host the MORPH controller processes. We make a note here that the optimality of controller placement decisions is orthogonal to the issue solved in this paper, and was thus not considered crucial for our evaluation. The resulting controller placement is depicted in Fig. \ref{fig:topo1}. The SDN controller replicas in the data-center topology were deployed on the leaf-nodes, similar to the controller placement presented in \cite{huang2017dynamic}. 

\begin{table}[htb]
	\centering
	\caption{Parameters used in evaluation of MORPH implementation}
	\begin{tabular}[htb]{ l l l l }
		  \hline
		  Parameter & Intensity & Unit & Meaning  \\
		  \hline
		  $F_M$ & $[1,2,3,4,5]$ & N/A & Max. no. of Byzantine failures\\
		  $F_A$ & $[1,2,3,4,5]$ & N/A & Max. no. of availability failures\\
		  $|C|$ & $[4,7,10,13,16]$ & N/A & No. of deployed controllers\\
		  $1/{\lambda_{F_M}}$ & $[5, 10, 15, 20]$ & [s] & Byzantine failure rate\\
		  $1/{\lambda_{F_A}}$ & $[5, 10, 15, 20]$ & [s] & Availability failure rate\\
		  $S$ & $[0,1]$ & N/A & SIA/SDA operation\\
		  $E$ & $[0,1]$  & N/A & NON-SELECTIVE/SELECTIVE\\
		  $T$ & $\emph{internet2, fat-tree}$  & N/A & Topology type\\
		  \hline
	\end{tabular}
	\label{table:params}
\end{table}

To evaluate the effect of the number, ratio and order of the Byzantine and availability-type failures, we deploy a specialized fault injection component. The fault injection component generates the faults in any of the available SDN controller processes using an out-of-band management channel. The targeted active controllers are faulted with uniform probability. The ratio of failure injections is manipulated by specifying the maximum amount of individual type of failures (ref. Table \ref{table:params}). The ordering of the different failure type injections is governed by the parametrization of mean arrival times as specified in Table \ref{table:params}. The  intensity of failures follows the negative exponential distribution (similar to \cite{nencioni2016availability, sakicresponse}).

The Docker- and OVS-based topology emulator, the REASSIGNER as well as up to \emph{16} C-COMPARATOR and \emph{34} S-COMPARATOR processes were deployed on a single commodity PC running a recent multi-core AMD Ryzen 1600 CPU and 32 GB of DDR4 RAM. We have used Gurobi Optimizer\footnote{Gurobi Optimizer - \url{http://www.gurobi.com/products/gurobi-optimizer}} to solve the ILP formulated in Sec. \ref{ilp}. 


\section{Results}
\label{results}

\subsection{Overhead minimization by successful failure detection} 

\begin{figure*}
	\centering
	\subfloat[Internet2: Controller cluster convergence time]{\includegraphics[width =0.5\textwidth]{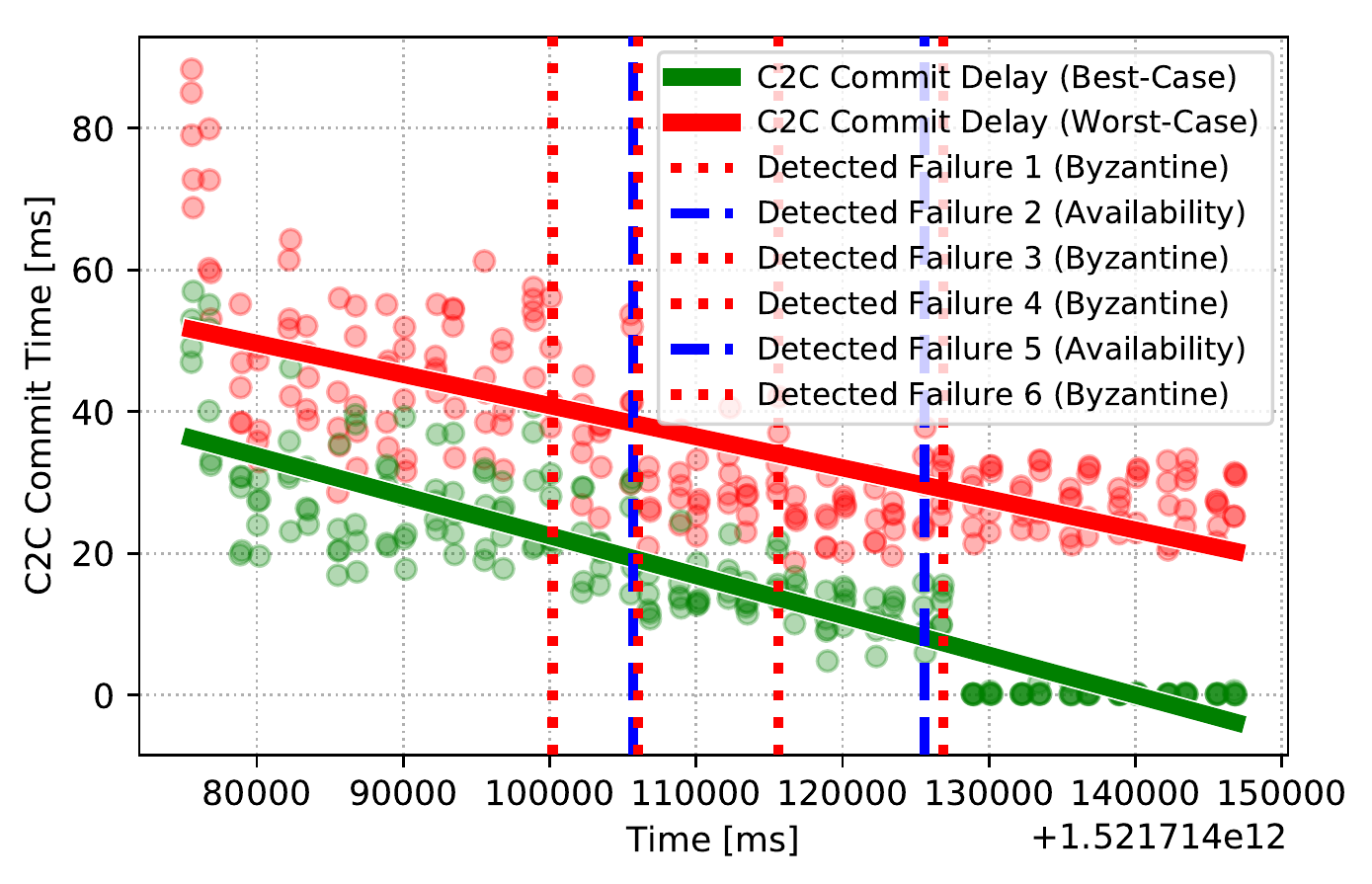}}
	\subfloat[Internet2: Switch reconfiguration delay]{\includegraphics[width = 0.53\textwidth]{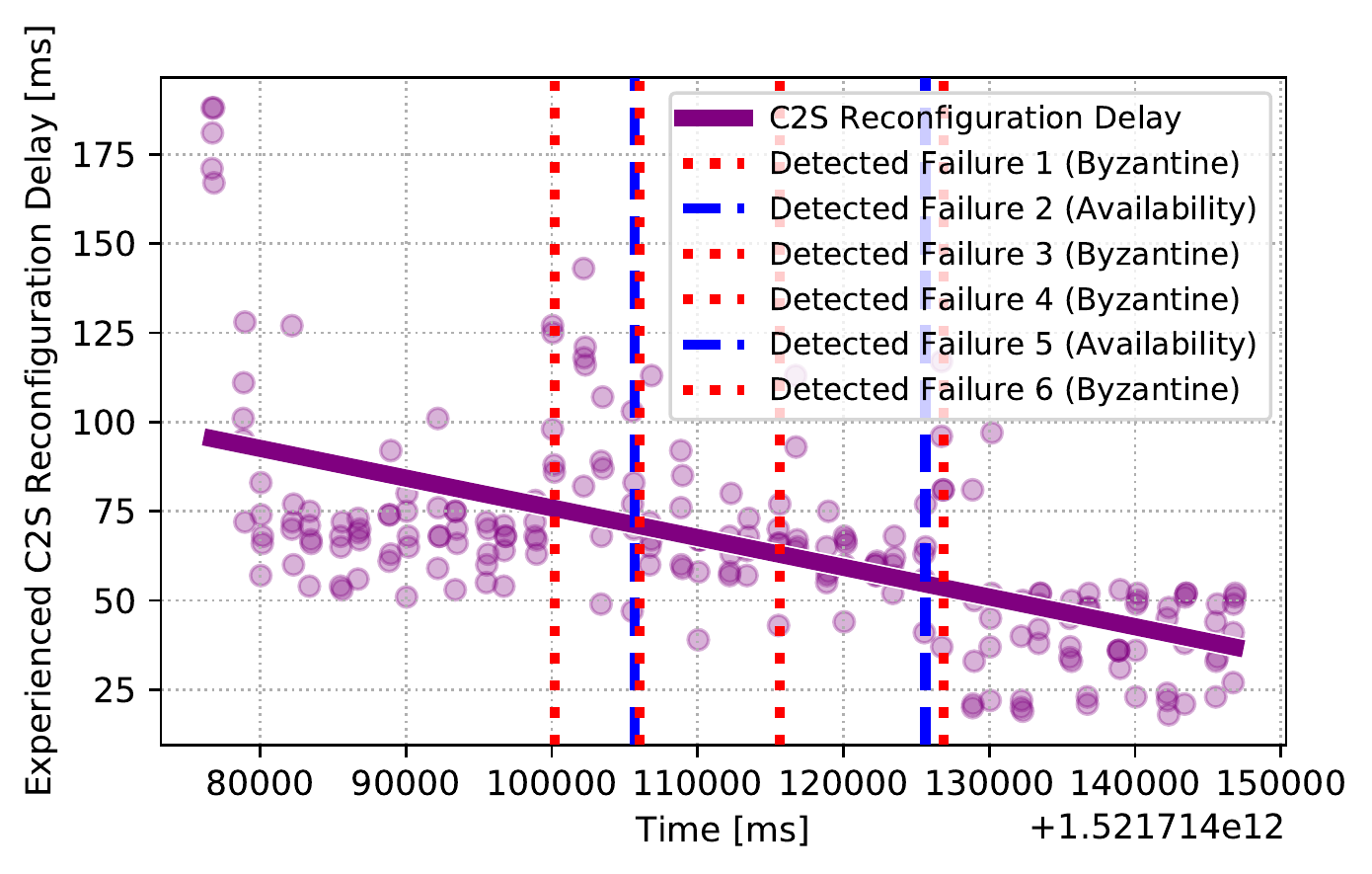}}\\
	\subfloat[Internet2: Observed received no. of packets over time]{\includegraphics[width = 0.5\textwidth]{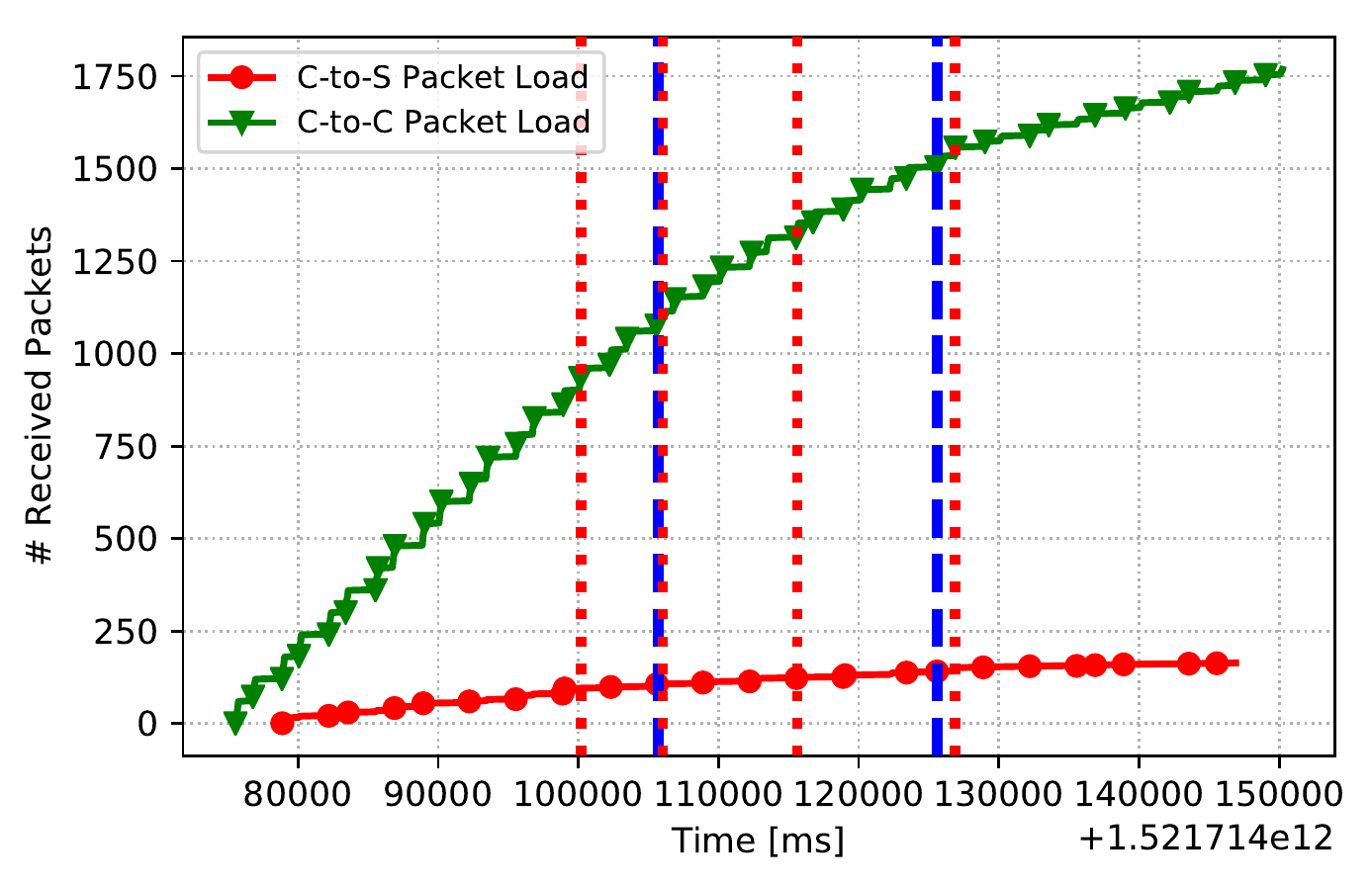}}
	\subfloat[Internet2: Observed CPU load of system over time]{\includegraphics[width = 0.51\textwidth]{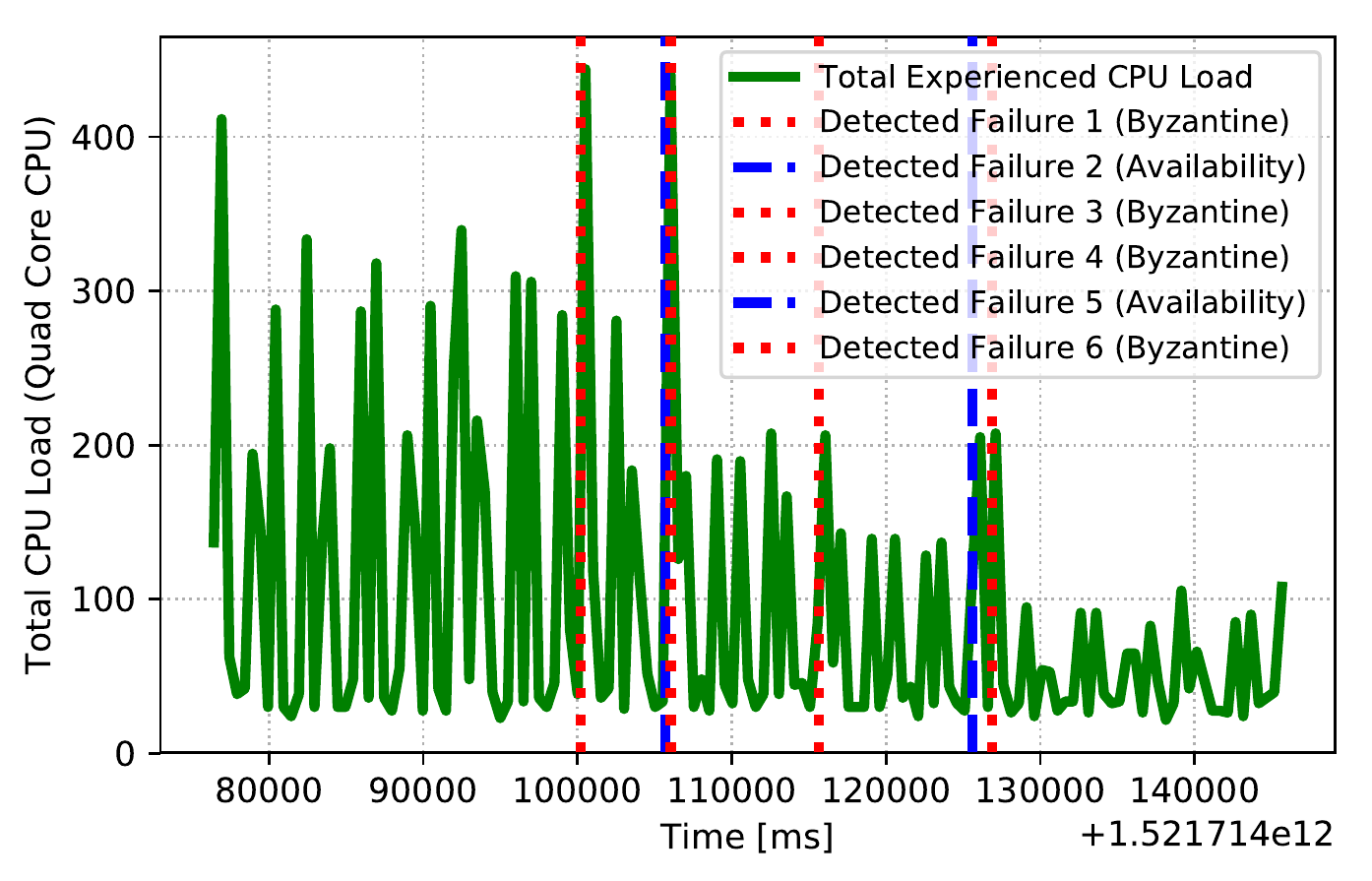}}
	\caption{Successful detection of injected Byzantine and availability failures leads to an instant decrease in: a) the experienced controller-to-controller (C2C) delay, both for the duration of the best- and worst-case state convergence time; b) the switch reconfiguration delay; c) the measured C2C and controller-to-switch (C2S) packet arrivals (measured at the ingress of a correct controller and switch, respectively); and d) the total measured system CPU load. The depicted measurements are taken in the Internet2 topology, based on $13$ controllers and $34$ switches. The particular sample shown here depicts the case where $4$ Byzantine and $2$ availability failures are identified in the depicted order (red and blue vertical dashed lines, respectively). }
	\label{fig:overheadminsingle}
\label{fig_1}
\end{figure*}

Fig. \ref{fig:overheadminsingle} depicts the observed controller-to-controller (C2C) state reconfiguration and switch-reconfiguration (C2S) delays, before, during and following the failure injection process. Similarly, the packet load and total system CPU utilization are depicted for the same measurement. Fig. \ref{fig:overheadminsingle} a) depicts the convergence time to the globally consistent state in the replicated controller database. The \emph{Best-Case} considers the time required to replicate and commit a state update in at least two controllers, while the \emph{Worst-Case} considers the outcome where \emph{every} controller converges to the globally consistent state. We observe that the initial failure injections consistently decrease the expected waiting time to achieve consensus and converge the state updates for the worst-case. The detection of later injections (Injection 5 and 6) has a lesser effect. For those detections, the trade-off in the added C2C delay overrides the benefits of the lower amount of controller confirmations required to update the controller state on each instance. This is a consequence of the displacement of the remaining active controllers in the Internet2 topology and hence the added controller-to-controller delay.

Fig. \ref{fig:overheadminsingle} b) depicts a drastic decrease in the response time required to handle client requests, i.e., to reconfigure each switch on the identified path with the new flow rules. We observe that \emph{dynamically decreasing the minimum number of controllers needed to reach consensus on the correct configuration update, consistently lowers the expected time to reconfigure both the control and data plane}. Thus, the overall throughput of the SDN control plane is increased, as additional service requests may be served in the same amount of time.

Similarly, the exclusion of incorrect controllers from the message comparison procedure consistently lowers the experienced control plane packet and CPU load over the observation period. The exclusion of the controller instances, specifically, has a benefit on the minimization of the number of C2C synchronization messages, as can be observed from the change in the arrival slope in Fig. \ref{fig:overheadminsingle} c). The total CPU load is similarly decreased with each successfully identified controller fault, as both controller and switches must process a lower total amount of controller messages with each controller exclusion. We have observed identical trends for the fat-tree topology.
 
Note that in static controller-to-switch assignment, the depicted downward slopes would be non-existent, i.e., a comparable design \cite{mohan2017primary, li2014byzantine} with static cluster configuration would not lead to a noticeable decrease in either CPU load or the switch/controller response times, even after the discovery of the faulty controller instances. This is in part because the mentioned comparable designs tend to ignore controller messages after detecting a Byzantine fault, instead of fully excluding the controllers from the actual cluster configuration.

\begin{figure*}
	\centering
	\subfloat[Internet2: Controller cluster convergence time]{\includegraphics[width =0.45\textwidth]{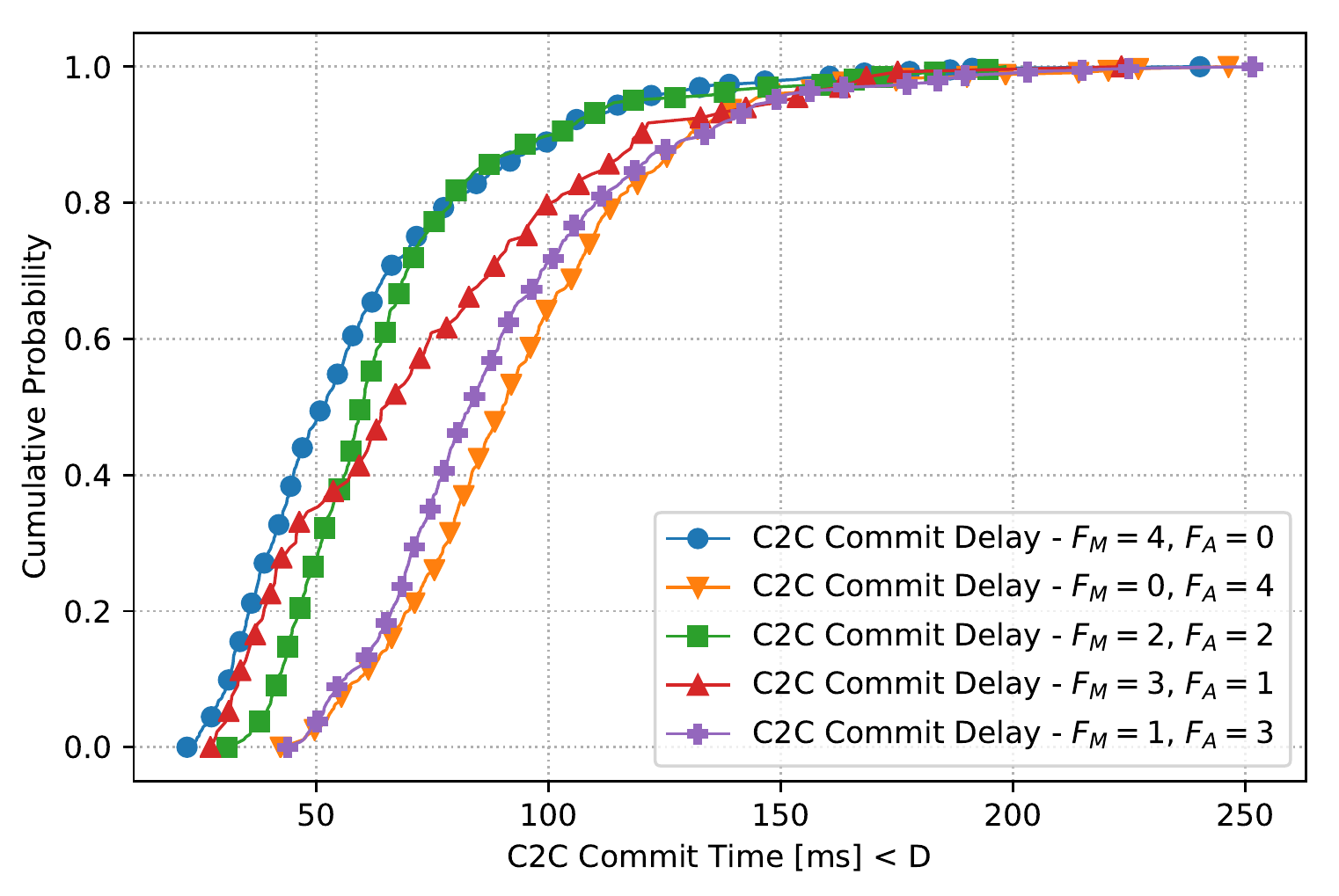}}
	\subfloat[Internet2: Switch reconfiguration delay]{\includegraphics[width =0.44\textwidth]{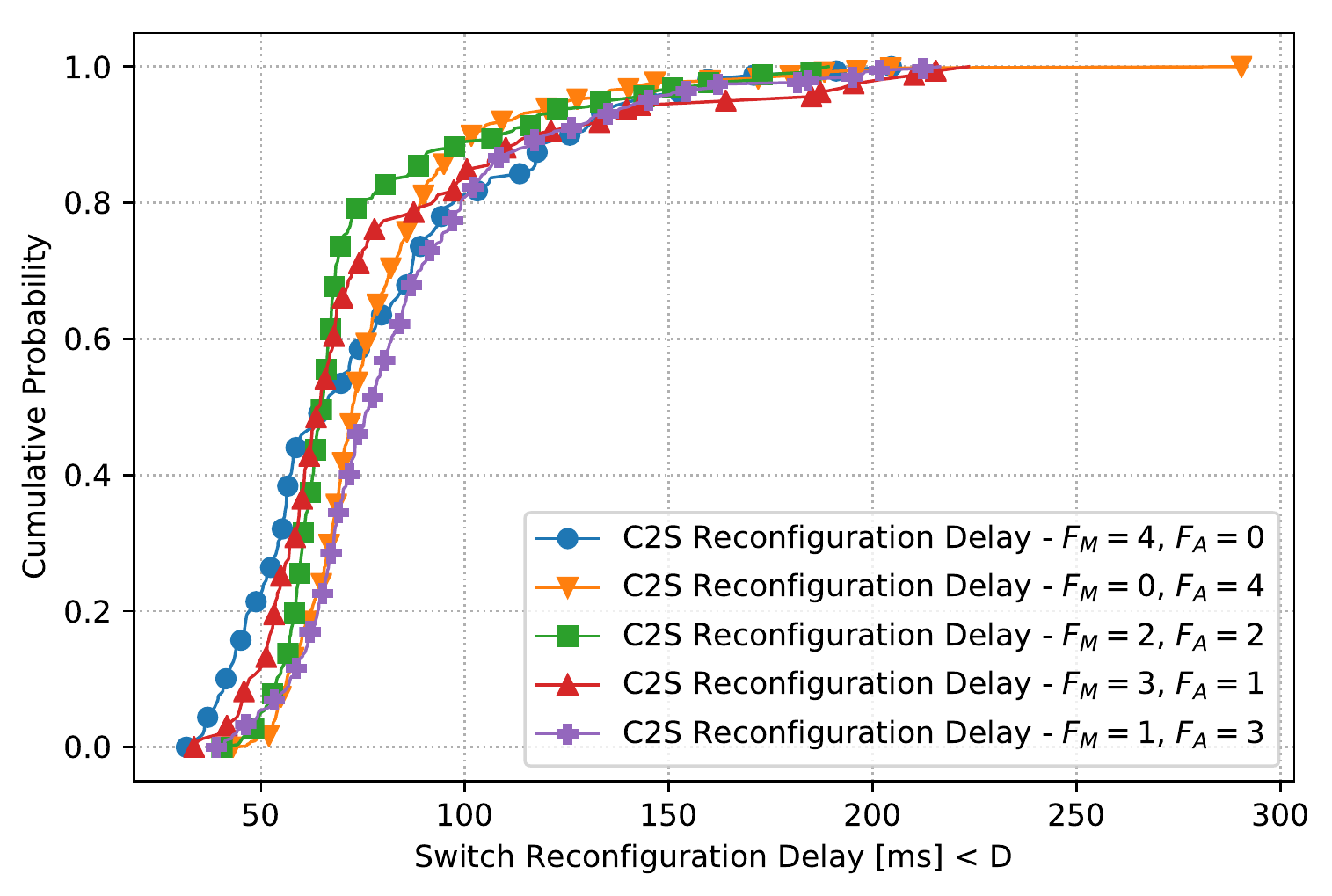}}\\
	\caption{The type of the failure affects considerably the experienced controller-to-controller delay (depicted in Fig. \ref{fig:overheaddecreasetypefailure} (a)). Indeed, the larger the number of detected Byzantine controller failures, the lower the instantaneous total number of PRIMARY controller messages required to process and commit new controller state update. Interestingly, the trend is observable for the switch configuration delay as well (depicted in Fig. \ref{fig:overheaddecreasetypefailure} (b)), but not with the same intensity. The depicted figures result from the measurements taken for the Internet2 topology comprised of $13$ controllers and $34$ switches. The observation period includes the time preceding, during and following the successful failure injections.}
\label{fig:overheaddecreasetypefailure}
\end{figure*}

Fig. \ref{fig:overheaddecreasetypefailure} depicts the benefits of distinguishing the type of failure when reducing the number of active controllers used in the consensus procedure. According to Alg. \ref{shuffle}, two message exchanges less are required to identify the correct message after detection of a malicious controller ($F_M$), while a single message less is required after discovery of an unavailable controller ($F_A$). Successful detection of Byzantine controllers thus leads to a lower average C2S and C2C reconfiguration time, compared to a discovery of the same amount of failures of a different type (or a combination of different faults). The detection of an availability failure ($F_A$) hence provides less information about the failed controller being either benign or malicious. In that case, the REASSIGNER is more conservative when computing new controller-switch assignments.

\subsection{Impact of SDN application statefulness} 
\begin{figure}
	\centering
	\subfloat[Internet2: Controller cluster convergence time]{\includegraphics[width =0.41\textwidth]{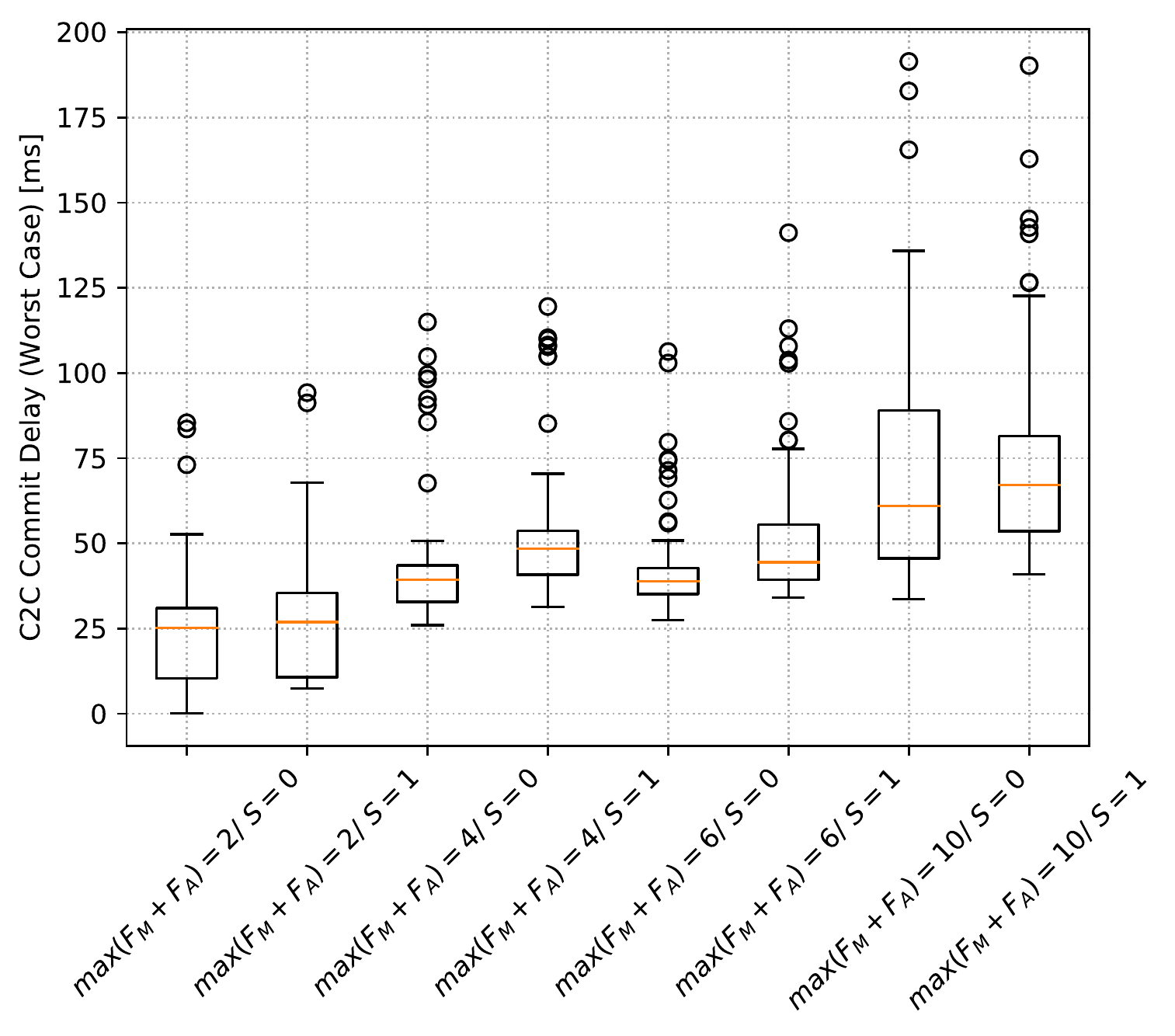}}\\
	\subfloat[Internet2: Switch Reconfiguration Delay]{\includegraphics[width =0.45\textwidth]{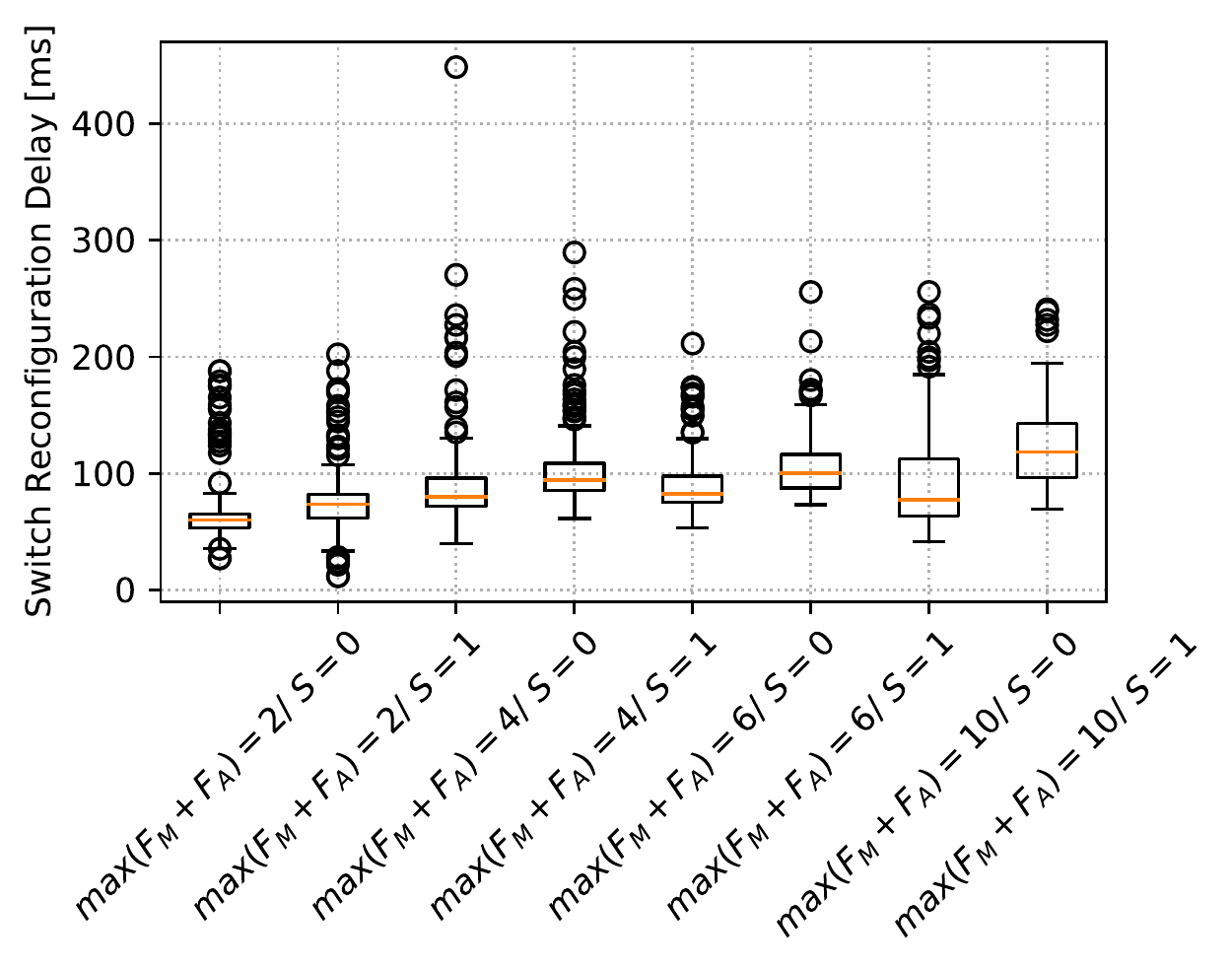}}\\
	\subfloat[Fat-Tree: Switch Reconfiguration Delay]{\includegraphics[width =0.45\textwidth]{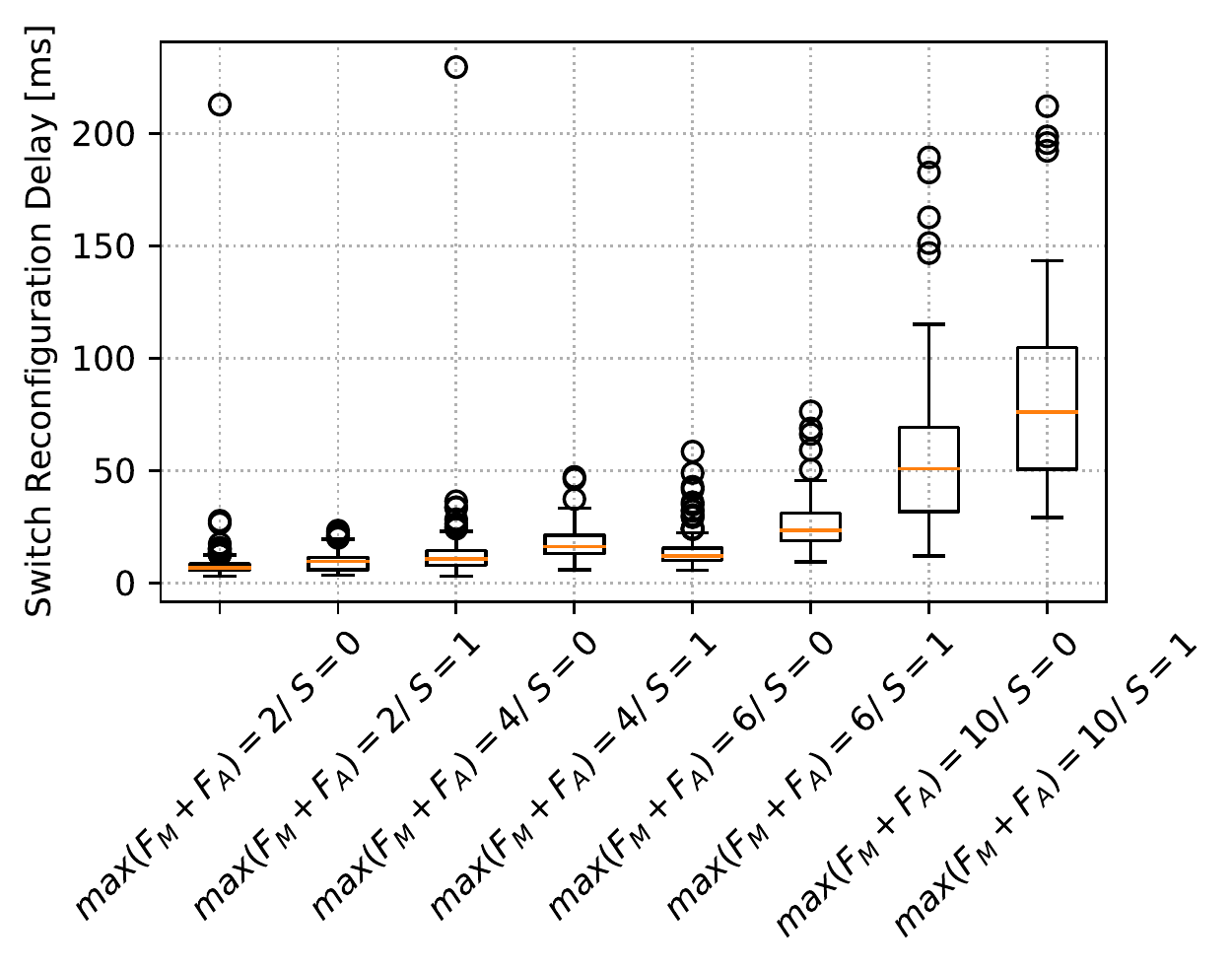}}
	\caption{Depending on an SDN application's requirement for access to the consistent controller data-store, controllers impose a varying controller-to-controller synchronization load. Specifically in the case of resource-reservation applications, knowledge about the reserved and free resources is a requirement for the efficient client request processing (i.e. slicing, path embedding, load-balancing algorithms that rely on resource reservations). $S=1$ denotes the case for a stateful controller application, while $S=0$ denotes applications which do not demand causality in controller updates.} 
\label{fig:i2_statefulness}
\end{figure}

Fig. \ref{fig:i2_statefulness} depicts the consequence of the complexity of statefulness of the distributed SDN control plane on the experienced C2C controller- and C2S switch-reconfiguration delays (Fig. \ref{fig:i2_statefulness} a) and Fig. \ref{fig:i2_statefulness} b), respectively). Stateful SDA applications necessitate consensus, which imposes an additional waiting time in the C2C synchronization. With consensus, the majority of controller configurations must match before deciding on the configuration message to be delivered to switch. After the consensus is reached, the controllers' data-stores converge to a consistent state in each correct controller replica. The difference in the waiting time is not reflected in the data plane (switch) as much as in the C2C communication, since the switch experiences a constant waiting time for the minimum amount of required replicas $Req_P$, independent of the time it takes to reach consensus in the C2C communication. 

However, as depicted both in Fig. \ref{fig:i2_statefulness} and Fig. \ref{fig:ftree_efficiency}, the overhead in the controller state and switch reconfiguration time is intensified by the \emph{maximum number} of tolerated failures. The more failures the system is designed to tolerate, the higher the amount of required comparisons across the PRIMARY messages in order to forward the system state. We conclude that \emph{the experienced controller and switch reconfiguration times scale with the number of tolerated controller failures}.

\subsection{Impact of proactive propagation of switch configurations} 

\begin{figure}
	\centering
	\includegraphics[width =0.45\textwidth]{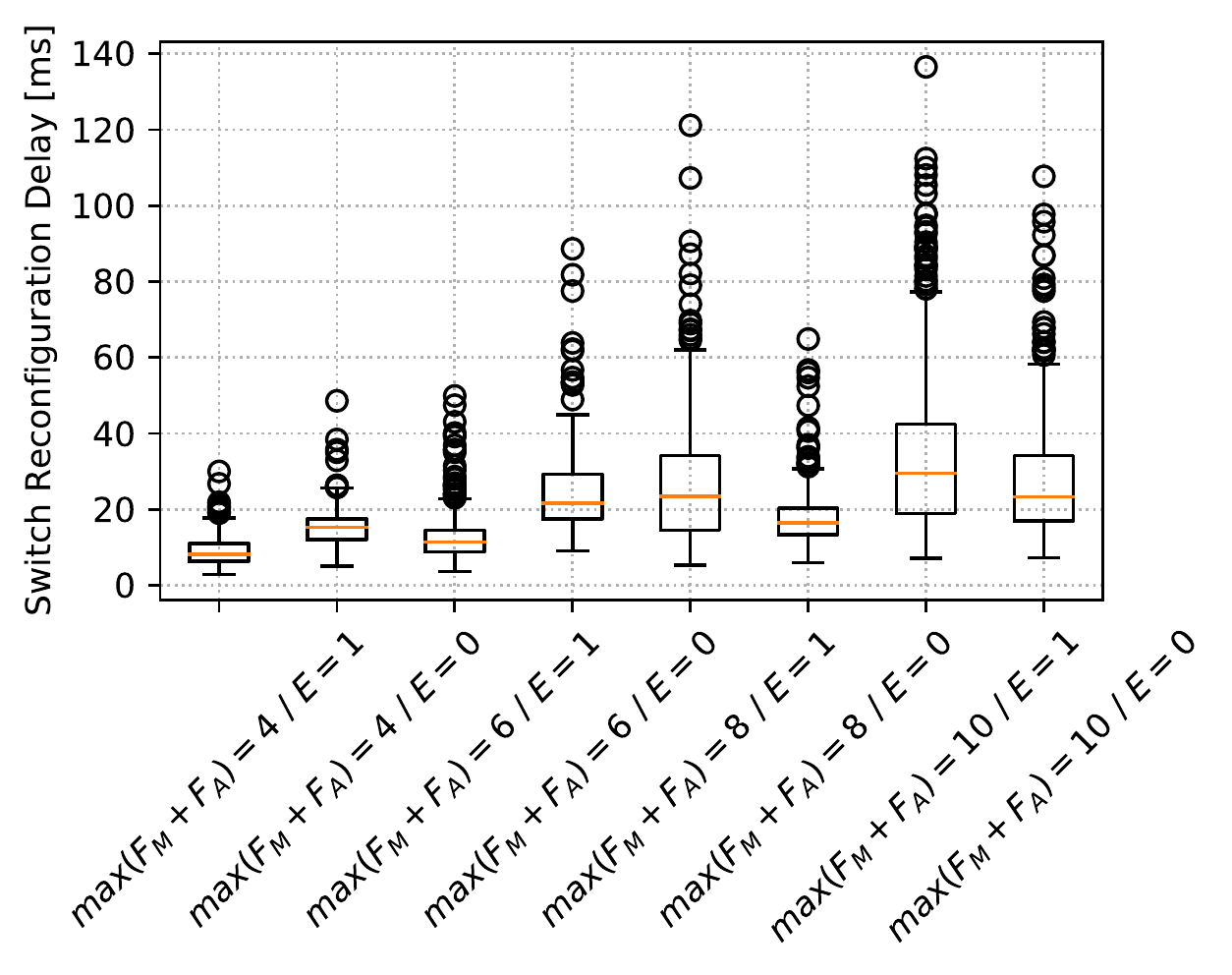}
	\caption{Difference in the observed switch reconfiguration delay for NON-SELECTIVE and SELECTIVE propagation of SECONDARY controller configurations in the $[7..16]$ controller topology that tolerates $max(F_M+F_A)=[4..10]$ controller failures. Deployment of a large control plane induces a benefit in the average configuration time with the NON-SELECTIVE model. In a topology comprising a smaller number of controllers, the additional system workload related with the propagation of configurations on every SECONDARY controller negatively affects the overall performance.}
\label{fig:ftree_efficiency}
\end{figure}

Fig. \ref{fig:ftree_efficiency} depicts the measurements taken in the fat-tree topology comprising 7 to 16 SDN controller instances and 20 switches. A trend reversal in the switch reconfiguration delay can be observed, where for small-sized controller clusters (i.e. $4..10$ controllers), the usage of the SELECTIVE mode ($E=1$) results in a lower configuration delay overall. For large-scale control planes ($13..16$ controllers), the trend is reversed and the NON-SELECTIVE mode ($E=0$) becomes faster. We assume that the trend reversal is linked to the fact that in the average case where no failures are detected, SELECTIVE mode is more efficient than the NON-SELECTIVE one, since no additional packet overhead / CPU load is generated during the message processing, both in the control and data plane. In the large-scale control plane, the controller placement has a dominant role, so that the decrease in the controller-switch distance dominates the additional load related to the higher number of controller instances. For client requests where some controllers send their malicious configurations to the switches, or fail-silently and thus do not deliver any new configuration messages, the switch is able to deduce the correct majority only after experiencing an additional round-trip to collect additional configuration responses from the SECONDARY controllers. The NON-SELECTIVE model is intuitively faster in its worst-case, as it propagates every controller response directly to the switch after the computation of the configuration response, independent of the controller's role as PRIMARY or SECONDARY. Thus, no additional switch-controller round-trip delays are experienced when the received messages at the switch are inconsistent. Similarly, the switch is capable of collecting the $Req_P$ consistent messages faster on average, since SECONDARY controllers may potentially deliver new configurations faster than the assigned PRIMARY instances. 

\subsection{Controller-switch reassignment time} 

\begin{figure}
	\centering
	\includegraphics[width =0.5\textwidth]{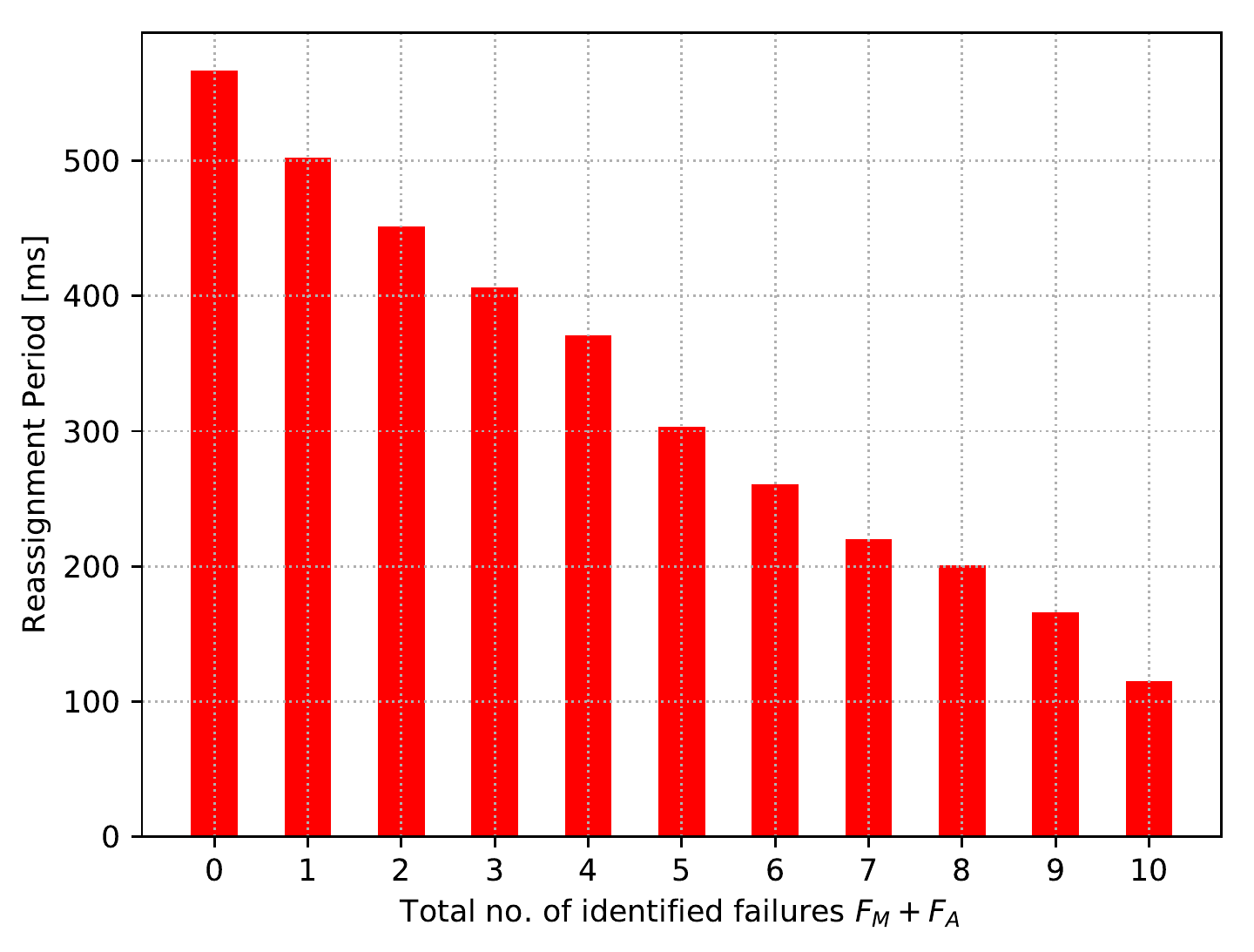}
	\caption{Decrease in the total execution time of the ILP solver for the controller-to-switch connection reassignment procedure. The controller-switch assigment (re-)computations consider $0-10$ failure detections and were executed in an emulated $34$-switch, $16$-controller Internet2 network topology. The time required for the REASSIGNER to reconfigure the control plane scales inversely with the number of successfully identified controller faults. As the REASSIGNER tasks may execute asynchronously, without negatively disturbing the system correctness or control plane availability, we consider the deployment of MORPH in critical operation networks for practically feasible.}
\label{fig:ilpdecreasetypefailure}
\end{figure}

The ILP solver is executed once after the initial system deployment (to compute the initial switch-controller assignments) and once after each controller failure detection. Fig. \ref{fig:ilpdecreasetypefailure} depicts a constant decrease in the execution time of the ILP as new faults are injected and failures are observed at the REASSIGNER. Indeed, the exclusion of incorrect controllers from the parameter space decreases the total running time of the ILP solver. Nevertheless, even in a large-scale formulation that assumes the assignment of $16$ controller replicas to $34$ switches, the REASSIGNER requires only up to $\sim540ms$ to compute the optimal assignment w.r.t. delay and bandwidth constraints. Hence, MORPH supports the real-time switch reassignment and is thus deployable in online operation as well. For higher-scaled networks we do not foresee any limitations in the reassignment procedure related to the viability of our solution. Namely, when an inconsistency in the PRIMARY configuration messages is detected in the S-COMPARATOR, the switch initially ignores the inconsistent messages and continues to autonomously contact the SECONDARY replicas to request additional confirmation messages. Thus, the correctness of the system is never endangered and the REASSIGNER may proceed with its reassignment optimization procedure asynchronously in the background. MORPH is hence suitable for operation in critical networks where minimal downtime is a necessary prerequisite (e.g., in industrial control networks).


\section{Conclusion}
\label{conclusion}

MORPH allows for distinguishing malicious/unreliable from unavailable controllers during runtime, as well as for their dynamic exclusion from the system configuration. In general, enabling Byzantine Fault Tolerance results in a relatively high system footprint in the network phase where no Byzantine or availability-related failures are identified. However, in the critical infrastructure networks, such as the industrial networks, this overhead may be unavoidable. 

Following a Byzantine or availability-induced SDN controller failure, MORPH allows for an autonomous adaptation of the system configuration and minimization of the distributed control plane overhead. By experimental validation, we have proven that MORPH achieves a performance improvement in both average- and worst-case system response time by minimizing the amount of required / considered controllers when deducing new configurations. Thus, the worst-case waiting periods required to confirm new controller state updates and switch configurations are substantially reduced over time. Furthermore, by excluding faulty controllers, the average packet and CPU loads incurred by the generation and transfer of controller messages to the switches are reduced. We have shown that the ILP formulation for QoS-constrained reassignment of controller-to-switch relationships may execute online, for both medium- and large-scale control planes comprising up to $16$ controller instances. The time required to re-adapt the system configuration scales proportionally with the number of successfully detected faulted controllers. Apart from the minimal additional CPU load related to the dynamic reassignment of controller connections, average- and worst-case computational and communication overheads are lower than those of comparable BFT SDN designs.

\section*{Acknowledgment}

This work has received funding from the EU's Horizon 2020 research and innovation programme under grant agreement number 780315 SEMIOTICS and CELTIC EUREKA project SENDATE-PLANETS (Project ID C2015/3-1) and is partly funded  by  the  BMBF  (Project  ID  16KIS0473). We are grateful to Hans-Peter Huth, Dr. Johannes Riedl, Dr. Andreas Zirkler, and the reviewers for their useful feedback and comments.


%
\bibliographystyle{IEEEtran}
\bibliography{IEEEabrv,qos}

%



\vspace{-1cm}
\begin{IEEEbiography}[{\includegraphics[width=0.9in,height=1.25in,clip,keepaspectratio]{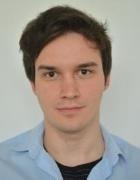}}]{Ermin Sakic} (S$'$17) received his B.Sc. and M.Sc. degrees in electrical engineering and information technology from Technical University of Munich in 2012 and 2014, respectively. He is currently with Siemens AG as a Research Scientist in the Corporate Technology research unit. Since 2016, he is pursuing the Ph.D. degree with the Department of Electrical and Computer Engineering at TUM. His research interests include reliable and scalable Software Defined Networks, distributed systems and efficient network and service management.
\end{IEEEbiography}
\vspace{-1.2cm}
\begin{IEEEbiography}[{\includegraphics[width=0.9in,height=1.25in,clip,keepaspectratio]{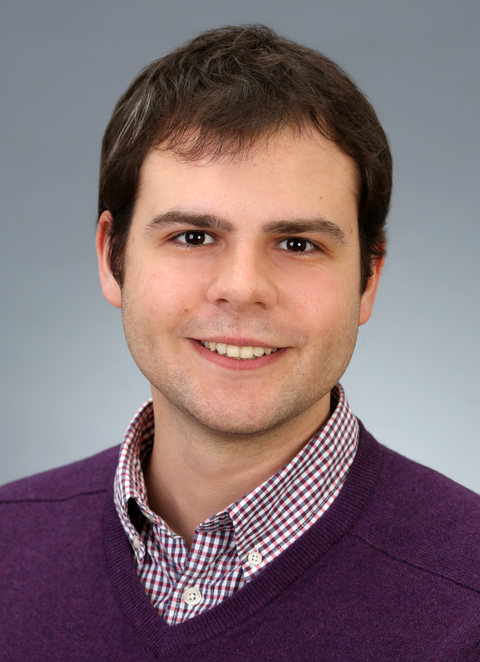}}]{Nemanja \DJ{}eri\'c{}} (S$'$18) received his B.Sc. degree in electrical and computer engineering in 2014, while he received M.Sc. degree in communication engineering from Technical University of Munich in 2016. He then joined the Chair of Communication networks at Technical University of Munich, where he is currently pursuing Ph.D. degree and working as a Research and Teaching Associate. His current research interests are Software Defined Networking (SDN) and Network Virtualization (NV).
\end{IEEEbiography}
\vspace{-1.2cm}
\begin{IEEEbiography}[{\includegraphics[width=1in,height=1.25in,clip,keepaspectratio]{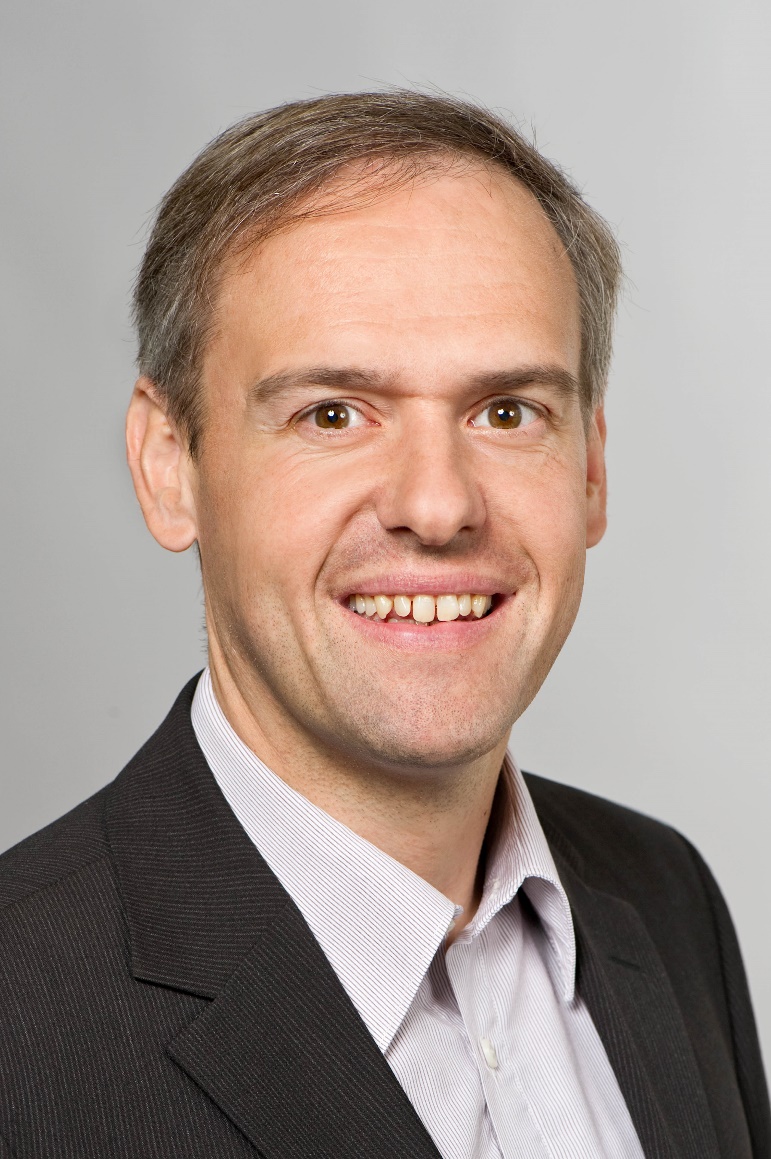}}]{Wolfgang Kellerer}
(M$'$96 \textendash\ SM$'$11) is a Full Professor with the Technical University of Munich (TUM), heading the Chair of Communication Networks at the Department of Electrical and Computer Engineering. Before, he was for over ten years with NTT DOCOMO's European Research Laboratories. He received his Dr.-Ing. degree (Ph.D.) and his Dipl.-Ing. degree (Master) from TUM, in 1995 and 2002, respectively. His research resulted in over 200 publications and 35 granted patents. He currently serves as an associate editor for IEEE Transactions on Network and Service Management and on the Editorial Board of the IEEE Communications Surveys and Tutorials. He is a member of ACM and the VDE ITG. \end{IEEEbiography}

\end{document}